\documentclass[12pt]{article}

\usepackage{newtxtext,newtxmath}
\usepackage{graphicx}

\usepackage{xurl}   % improves URL line breaking
\usepackage{url}    % usually already loaded, but safe
\usepackage{textcomp}
\usepackage{graphicx}
\usepackage{hyperref}
\usepackage{float}
\usepackage[T1]{fontenc}
\usepackage{textgreek}
\usepackage{longtable}
\usepackage{array}
\usepackage{caption}
\usepackage{physics}
\usepackage{booktabs}

\usepackage[letterpaper,margin=1in]{geometry}

\renewenvironment{abstract}
	{\quotation}
	{\endquotation}

\date{}

\makeatletter
\renewcommand{\fnum@figure}{\textbf{Figure \thefigure}}
\renewcommand{\fnum@table}{\textbf{Table \thetable}}
\makeatother

\usepackage{scicite}

\usepackage{url}

\def\scititle{
	Sub-Doppler rubidium atom cooling using a programmable agile integrated PZT-on-SiN resonator
}
\title{\bfseries \boldmath \scititle}

\usepackage{authblk}

\author[1$\dagger$]{Andrei Isichenko}
\author[2$\dagger$]{Steven Carpenter}
\author[1]{Nick Montifiore}
\author[1]{Jiawei Wang}
\author[4]{Mayand Dangi}
\author[1]{Nitesh Chauhan}
\author[4]{Pritha Mukherjee}
\author[2]{Xuting Yang}
\author[1]{Nitin Indukuri}
\author[1]{Mark W. Harrington}
\author[1]{Chuan Zhong}
\author[3]{Iain M. Kierzewski}
\author[3]{Ryan Q. Rudy}
\author[2,4$^\ast$]{Jennifer T. Choy}
\author[1$^\ast$]{Daniel J. Blumenthal}

\affil[1]{Department of Electrical and Computer Engineering, University of California Santa Barbara, Santa Barbara, California 93106, USA}
\affil[2]{Department of Physics, University of Wisconsin-Madison, Madison, Wisconsin 53706, USA}
\affil[3]{U.S. Army Research Laboratory, Adelphi, Maryland 20783, USA}
\affil[4]{Department of Electrical and Computer Engineering, University of Wisconsin-Madison, Madison, Wisconsin 53706, USA}
\affil[$\dagger$]{These authors contributed equally to this work.}
\affil[$\ast$]{Corresponding Authors: jennifer.choy@wisc.edu, danb@ucsb.edu}
\begin{document} 

% Insert the title and author list
\maketitle
% Abstract, in bold
% There are strict length limits, and not all formats have abstracts.
% Consult the journal instructions to authors for details.
% Do not cite any references in the abstract.
\begin{abstract} %\bfseries \boldmath
% The abstract should be a single paragraph (less than 160 words) written in plain language that a general reader can understand. Do not include citations. The abstract should provide:
% • An opening sentence that states the question/problem addressed by the research
% • Enough background content to give context to the study
% • A brief statement of primary results
% • A short concluding sentence
\textbf{Abstract:} Programmability and precise control of laser frequency are essential for quantum experiments and applications such as atomic clocks, quantum computers, and cold-atom sensors. Current systems use bulky, power-hungry modulators and frequency shifters which are difficult to integrate and limit portability and scalability. We report an electrically controllable, agile optical frequency source based on a semiconductor laser stabilized to a photonic-integrated, lead zirconate titanate (PZT)–actuated resonator cavity. We demonstrate this approach with precision programmable frequency control of a 780-nm laser that can periodically reference to rubidium spectroscopy followed by fast, programmable, arbitrary frequency tuning sequences for quantum control. We use this approach to demonstrate sub-Doppler cooling of rubidium-87 without any external modulators, achieving atom-cloud temperatures as low as 16 $\mu$K. The device achieves a tuning strength up to 1 GHz/V with 11 MHz modulation bandwidth while consuming only 10 nW of electrical power. This work establishes a route toward compact, low-power, and chip-scale laser systems for next-generation quantum and atomic sensing technologies.
\end{abstract}

\noindent

\subsection*{Introduction}

Precision laser frequency control underpins a wide range of quantum technologies, including optical atomic clocks \cite{ludlow_optical_2015}, quantum computing \cite{graham_multi-qubit_2022}, and cold-atom inertial sensors \cite{stray_quantum_2022, lee_compact_2022}. These systems rely on narrow-linewidth lasers that must resolve and be stabilized with respect to atomic transitions. For example, rubidium (Rb) systems operating near 780 nm require absolute frequency drift below the 100 kHz level for precision sensor performance \cite{loh_microresonator_2016, liang_compact_2015, theron_narrow_2015}. In a cold-atom magneto-optical trap (MOT), this frequency is red-detuned by tens of megahertz for trapping and then rapidly returned to resonance for fluorescence detection.  In addition, achieving sub-Doppler temperatures requires dynamic, precisely timed detuning sequences: polarization-gradient cooling (PGC) \cite{dalibard_laser_1989} and $\Lambda$-enhanced gray molasses \cite{rosi_-enhanced_2018} depend on rapid, millisecond-speed frequency ramps of tens of megahertz, while Raman transitions used for coherent state manipulation often demand gigahertz-scale frequency jumps. The ability to control these frequency offsets with both long-term stability and short-term agility is therefore fundamental to reaching the ultra-low temperatures and coherence needed for high-precision sensing and quantum-state readout.

Techniques for controlling laser frequency have achieved a high degree of maturity, with ongoing efforts to miniaturize and scale. Long-term laser frequency stability is conventionally achieved by referencing to atomic vapor-cell spectroscopy, such as saturation-absorption spectroscopy (SAS), while fast, programmable detuning is realized through external modulation or direct laser control (e.g., by tuning laser current and temperature) in combination with offset-beatnote locking \cite{wiegand_single-laser_2019}. Acousto-optic modulators (AOMs) can frequency shift by hundreds of megahertz, and electro-optic modulators (EOMs) apply sidebands that extend this range to multi-gigahertz offsets. These tools form the basis of many compact cold-atom instruments: single-seed laser architectures employing AOMs and EOMs have enabled field-deployable atom interferometers \cite{chiow_compact_2018} and recent photonic implementations at the C-band have demonstrated high-speed, single-sideband frequency shifting and modulation \cite{lee_compact_2022,kodigala_high-performance_2024}. However, these approaches achieve frequency agility at the cost of system complexity and power consumption. In particular, bulk-optic and fiber-coupled AOMs require watt-level RF power amplifiers and lose optical power to unwanted diffracted orders, while EOMs generate parasitic sidebands that can degrade atom interferometer sensitivity \cite{theron_narrow_2015, wang_agile_2022}. Integrated solutions operating at C-band must also rely on frequency doubling to reach visible wavelengths, introducing conversion inefficiency and packaging overhead. Offset beat-note locking of the frequency-agile cooling laser is a reliable solution, but requires extra lasers for operation \cite{wang_agile_2022}. Together, these limitations constrain the scalability of conventional architectures toward photonic-integrated, compact, low-power, field-deployable quantum systems.

A critical next step is the development of visible and near-infrared (NIR) precision frequency control at atomic transition wavelengths, such as the 780 nm $\mathrm{D}_2$ line of rubidium. Frequency modulation realized in the CMOS-compatible silicon nitride (Si$_3$N$_4$) waveguide platform is especially important for atomic applications due to its wide transparency window (405 nm – 2350 nm) \cite{blumenthal_silicon_2018}, ultra-low losses for laser frequency stabilization \cite{puckett_422_2021}, and compatibility with heterogeneous or hybrid integration with active materials such as GaAs, GaN, and InP. In particular, stress-optic modulation using lead-zirconate-titanate (PZT) on SiN is well suited to atomic frequency control  as it enables wideband (DC-coupled to tens of MHz bandwidth) and low-power control. PZT integration on silicon nitride introduces negligible additional optical loss and operates with nanowatt-level power consumption \cite{alexander_nanophotonic_2018, jin_piezoelectrically_2018, stanfield_cmos-compatible_2019, wang_silicon_2022, wang_integrated_2023}. PZT provides significantly stronger electromechanical coupling enabling larger frequency shifts at lower drive voltages. To date, most PZT modulation is in the 1550 nm waveband while recent work extending stress-optic modulation into the visible spectrum \cite{siddharth_narrow-linewidth_2025, montifiore_integrated_2025} has been reported. However, agile and arbitrary frequency control using integrated stress-optic tuning in the visible-to–shortwave infrared (SWIR) range has not yet been demonstrated or applied in an atomic experiment.

In this work, we demonstrate a 780 nm PZT-on-SiN stress-optic ring resonator that functions as a frequency-agile programmable optical cavity for laser frequency control and apply it to achieve sub-Doppler cooling of rubidium-87 atoms without any external modulation. Programmed voltage waveforms applied to the PZT actuator enable deterministic and repeatable control of the Pound-Drever-Hall (PDH)-locked laser frequency. We measure a static tuning efficiency of 1 GHz/V, 11-MHz small-signal modulation bandwidth, and sub-100 nW electrical power consumption, enabling precise DC tuning and agile modulation without degrading the optical Q-factor or requiring additional optical power. Using this frequency-programmable cavity, we realize rapid optical frequency ramp sequences for polarization-gradient cooling (PGC) in a 3D magneto-optical trap (MOT), achieving atomic temperatures as low as 16 $\mu$K—well below the 146-$\mu$K Doppler limit—using only the PZT resonator and a semiconductor optical amplifier (SOA), with no AOM in the loop for frequency shifting or intensity control. The PZT cavity provides holdover between periodic re-referencing to rubidium spectroscopy, illustrating a compact, low-power route to laser frequency synthesis for cold-atom control. This approach is inherently compatible with self-injection locking, establishing a pathway toward fully integrated, frequency-programmable laser systems for quantum sensing and atom interferometry.

\subsection*{Results}

\begin{figure} % Do NOT use \begin{figure*}
	\centering
	\includegraphics[width=1.0\textwidth]{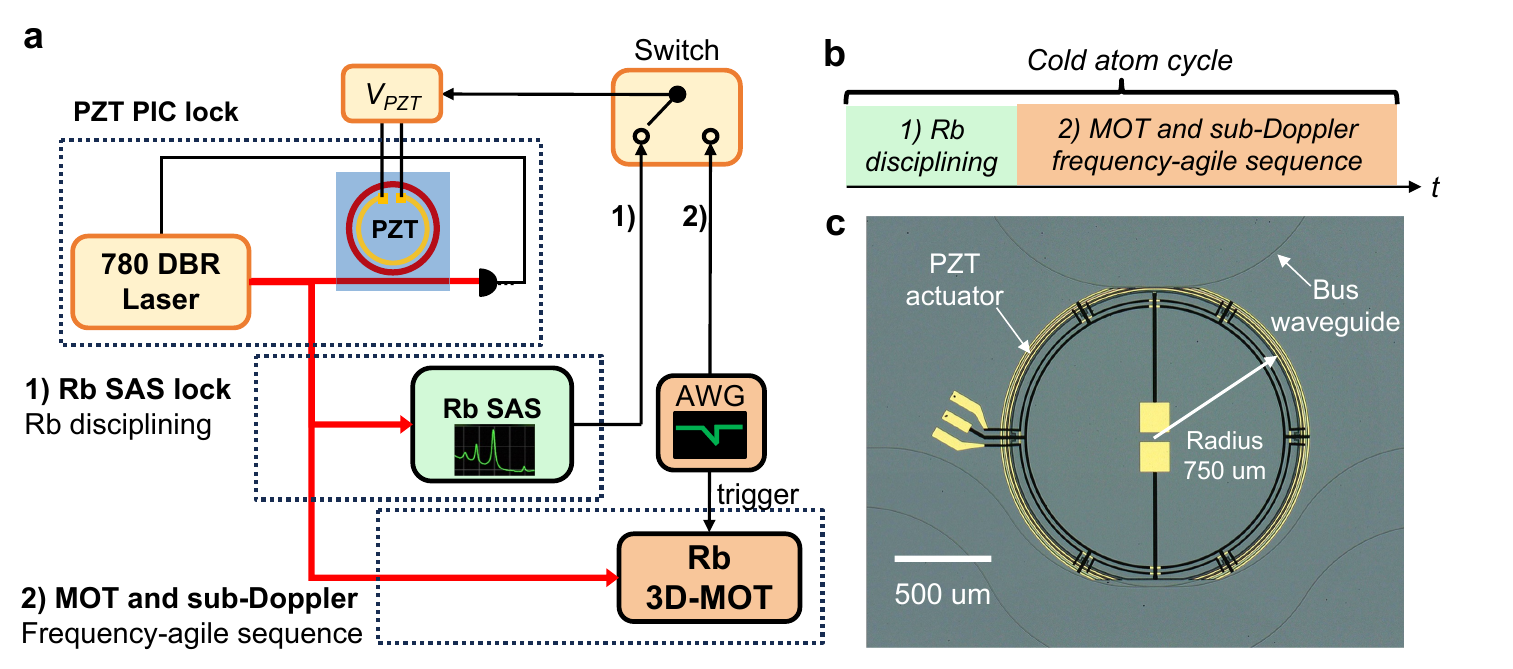}
	% Pick an appropriate width - in print, figures are usually one or two columns wide, which can
	% be approximated by 0.3\textwidth or 0.6\textwidth respectively. Use appropriate label sizes.
	\caption{\textbf{Working principle of rubidium sub-Doppler cooling using an agile integrated PZT-controlled silicon nitride cavity.}
		  \textbf{a)} A 780-nm laser is Pound-Drever-Hall (PDH) locked to an integrated resonator containing a PZT stress-optic actuator. The laser closely tracks the resonator (PZT PIC lock) and the laser drift is reduced at short timescales (i.e. drift holdover). The PZT is actuated such that the locked laser is stabilized to a rubidium hyperfine transition using saturation absorption spectroscopy (SAS), which provides Rb disciplining and long-term frequency stability. During the cooling cycle, we switch off the Rb lock (while PZT PIC lock remains active) and apply a fast control signal to the PZT to rapidly shift the laser frequency for the Rb MOT (magneto-optical trap) and for sub-Doppler atom cooling. \textbf{b)} The entire cold atom cycle consists of times when the laser and PZT PIC are Rb disciplined (1) and when the PZT PIC provides frequency agility and short-term holdover during the atom cooling experiment (2). \textbf{c)} Micrograph of the PZT-actuated SiN ring resonator cavity, ring radius 750 $\mu$m. 
}
	\label{fig:working_principle} 
\end{figure}

The PZT-on-SiN resonator provides an integrated platform for locked laser frequency control, combining short-term stability with fast tuning for programmable control and long-term atomic referencing. The working principle of the PZT-controlled agile resonator cavity is shown in Figure \ref{fig:working_principle}. A commercial single-frequency 780 nm semiconductor laser (Photodigm distributed Bragg reflector (DBR)) is Pound-Drever-Hall (PDH) locked to the resonator (Fig. \ref{fig:working_principle}A). This enables the locked laser to frequency-track the agile cavity while reducing the short-term laser frequency drift. During the laser stabilization step, we implement a dual-stage lock \cite{liang_compact_2015, isichenko_multi-laser_2025, loh_microresonator_2016} to a rubidium saturation absorption spectroscopy (SAS) module, where we feed the signal from the SAS back into the PZT tuner, anchoring the PIC-locked laser to an atomic transition and reducing the long-term laser frequency drift. 

To perform the detuning ramp sequences necessary for sub-Doppler atom cooling, the Rb spectroscopy lock is periodically disabled and an arbitrary waveform generator (AWG) drives the PZT with a pre-programmed voltage signal. This produces laser frequency ramps and jumps used for MOT formation and polarization gradient cooling (PGC). The process is repeated, periodically re-referencing to the Rb spectroscopy to prevent any significant long-term frequency drifts, while the PIC resonator serves as a holdover cavity to bound the frequency drift between cycles (Fig. \ref{fig:working_principle}B. The mm-scale PZT-on-SiN resonator device is shown in Fig. \ref{fig:working_principle}C.

\subsubsection*{Resonator characterization}

\begin{figure} 
	\centering
	\includegraphics[width=1.0\textwidth]{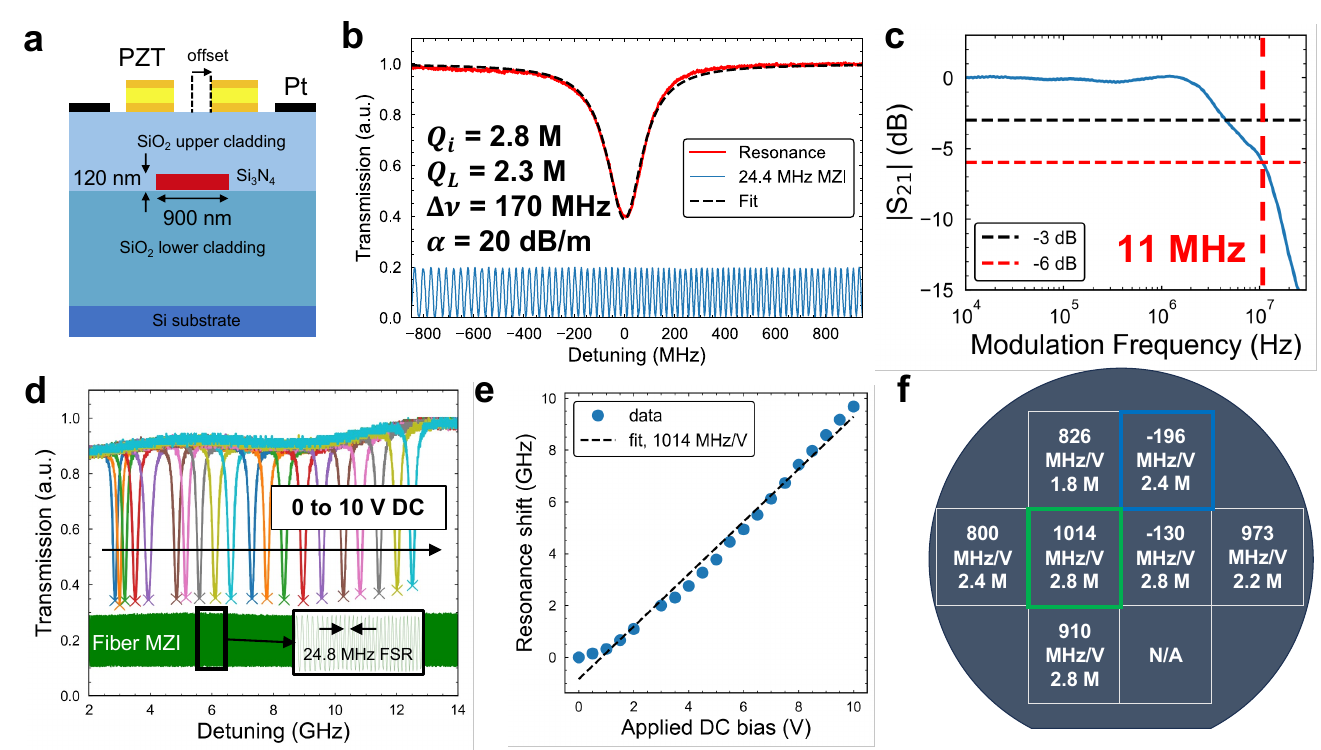}
	\caption{\textbf{PZT PIC resonator design and characterization.}
		  \textbf{a)} Cross-section of the silicon nitride ring resonator with PZT actuator and Pt thermal tuner, where the offset represents the lateral displacement between the PZT and the waveguide. \textbf{b)} Transmission spectrum quality factor (Q) measurement of the ring resonator results in an intrinsic quality factor ($Q_i$) of 2.8 million, a loaded quality factor ($Q_L$) of 2.3 M, loss $\alpha$=20 dB/m, and a total linewidth $\Delta \nu$ of 170 MHz. The frequency detuning is calibrated using an unbalanced Mach-Zehnder interferometer (MZI, blue trace). \textbf{c)} Frequency response of the PZT stress-optic small-signal modulation. The 6-dB modulation bandwidth is 11 MHz. \textbf{d)} Transmission spectra for a single resonance as a function of applied bias voltage to the PZT. The frequency tuning is calibrated with an unbalanced Mach-Zehnder interferometer (Fiber MZI, green trace) \textbf{e)} The resonance frequency shift as a function of the applied DC bias. The measured tuning strength is 1 GHz/V $\approx$ 2 pm/V. \textbf{f)} Map of the 4” wafer DUV stepper lithography exposure reticles. Each reticle contains many devices; we report the data (PZT tuning strength, Qi) for only a single add-thru resonator device in the reticle. Green outline: device with largest tuning strength, blue outline: device used for atom cooling demonstration, N/A: not applicable due to device damage.
}
	\label{fig:res_characterization}
\end{figure}

The cross-section design of the ring resonator is shown in Fig. \ref{fig:res_characterization}A. The waveguide thickness is chosen for a moderately confined mode for a smaller bend radius, and the width is designed to support the TM$_0$ mode. The resonator has a radius of 750 $\mu$m, corresponding to a 38-GHz free-spectral range (FSR). The layer stack consists of a 15-$\mu$m SiO$_2$ lower cladding, a 120-nm thick and 900-nm wide Si$_3$N$_4$ core, and a 4-$\mu$m SiO$_2$ upper cladding. The planar layer of PZT of 1-$\mu$m thickness is deposited on top of the waveguide with a lateral offset from the waveguide center. A pair of platinum electrodes is included for thermal tuning of the resonance. The PZT actuator deposition process does not require an undercut, unlike previous work \cite{jin_piezoelectrically_2018,stanfield_cmos-compatible_2019}. The details of the fabrication process are in the Materials and Methods section. 

The combination of the high quality factor, fast modulation, and large-range PZT tuning is central to the device performance. For the TM-mode resonance we measure an intrinsic $Q_i$ = 2.8 M, loaded $Q_L$ = 2.3 M, and a propagation loss of 20 dB/m (Fig. \ref{fig:res_characterization}B). We measure the small-signal electrical-to-optical modulation response $S_{21}$ by tuning the laser onto the quadrature point of the resonance and using a network analyzer. We find the 6-dB cutoff frequency to be 11 MHz (Fig. \ref{fig:res_characterization}C) and the 180$^{\circ}$ phase-lag point is 5.8 MHz (Fig.~\ref{fig:sup_PZT_AC_hyst}A).  The PZT actuator enables low-power frequency tuning over a broad range. The optical transmission spectrum (Fig. \ref{fig:res_characterization}D) shows resonance tuning as a function of DC control voltage from 0 to 10 V and a resonance shift corresponding to a linear tuning strength of 1 GHz/V (Fig. \ref{fig:res_characterization}E). Due to hysteresis in the stress-optic tuning, forward and reverse voltage sweeps yield slightly different resonance shifts, as shown in Fig.~\ref{fig:sup_PZT_AC_hyst}B. We measure the leakage current to be $<$1 nA at 10 V applied voltage, indicating a power consumption of $<$10 nW. To evaluate wafer-scale uniformity, we measure the static tuning and intrinsic $Q_i$ for devices across a 4" wafer (Fig. \ref{fig:res_characterization}F). All seven measured devices exhibit tuning strengths over 100 MHz/V, with the tuning polarity determined by the lateral offset of the PZT actuator relative to the waveguide, where positive (negative) tuning corresponds to no (2-$\mu$m) offset. The device from the green-outlined reticle achieved the highest $Q_L$ and tuning strength. The device from the blue-outlined reticle was packaged (Fig. \ref{fig:PZT_packaging}) and used for the atom cooling demonstration.

\subsubsection*{Sub-Doppler atom cooling}

We next demonstrate the application of the PZT-on-SiN resonator as an agile frequency tuner for a PDH-locked laser used in a rubidium cold-atom ensemble. This enables cooling of atoms in a three-dimensional magneto-optical trap (MOT) to temperatures well below the 146 $\mu$K Doppler limit through polarization-gradient cooling (PGC) \cite{dalibard_laser_1989}. Optimal PGC requires the MOT cooling light to be rapidly red-detuned, typically by 30–100 MHz within several milliseconds, while simultaneously reducing the optical power \cite{lee_compact_2022}. This dynamic detuning decreases the scattering rate and allows atoms to localize in the spatially varying light field, achieving sub-Doppler temperatures.

In this work, we focus on using the integrated resonator for frequency control. Conventionally, sub-Doppler cooling sequences rely on an acousto-optic modulator (AOM) for two functions: (1) shifting the laser frequency to the desired detuning for MOT formation, and (2) applying a chirped RF drive to ramp up the detuning during the sub-Doppler stage. AOMs can also act as optical shutters, although this role can alternatively be fulfilled by a semiconductor optical amplifier (SOA) or eventually by an extension of the same PZT mechanism presented in this work.

Using the agile frequency control of our PZT optical resonator, we replace the function of the AOM for both MOT formation and sub-Doppler cooling. In fact, no AOM is present in the path of the cooling beam. We control the system via a Vescent D2-125-PL servo controller which enables rapid switching between spectroscopy-referenced operation and free-running, resonator-stabilized operation. In the free-running mode, arbitrary voltage waveforms applied to the PZT allow rapid frequency control of the cooling laser.

The schematic used to demonstrate and measure the locked-laser tuning is shown in Fig. \ref{fig:exp_diagram}A. We PHD-stabilize the 780-nm cooling DBR laser to the ring resonator and achieve locked laser frequency control with $V_{\rm PZT}$. The applied PZT signal alternates between the ($V_{\rm Rb ~ lock}$) for Rb-disciplining and the fast frequency control ($V_{\rm agile}$) for MOT formation and for sub-Doppler cooling. The $V_{\rm agile}$ signal tunes the cooling laser frequency enabling arbitrary laser frequency control within the tuning range of the PZT actuator and at speeds within the locking bandwidth of the laser PDH lock as shown in Fig. \ref{fig:exp_diagram}B. A detailed schematic of the lock is shown in Figure \ref{fig:PZT_laser_locking} and discussed in Supplementary Materials section S1. 

We amplify the cooling light by a fiber-coupled semiconductor optical amplifier (SOA), whose output power is voltage-controlled ($V_{\mathrm{SOA}}$) for programmable power ramping and shuttering. The SOA serves as a power modulator for PGC and as an optical shutter during time-of-flight (TOF) temperature measurements. The $^{87}$Rb MOT employs a six-beam (three beams with retro-reflection) configuration. The 780-nm laser drives a red-detuned $\mathrm{^{87}Rb}~5S_{1/2}~F=2 \rightarrow 5P_{3/2}~F^\prime =3$ transition, while a separate 795-nm DBR laser acts as the repump, addressing the $\mathrm{^{87}Rb}~5S_{1/2}~F=1 \rightarrow 5P_{1/2}~F^\prime=2$ transition.

The timing diagram for one experimental cycle is shown in Fig.~\ref{fig:exp_diagram}C. Each sequence begins with Rb disciplining at the $F^\prime =(2,3)$ crossover transition for 25 ms. The PZT voltage is then stepped by $\sim$1 V to set the cooling laser at the MOT detuning $\Delta_{\mathrm{MOT}}$ relative to the $F^\prime =3$ cooling transition, and the MOT is loaded for 223 ms. Following MOT loading, the $V_{\mathrm{agile}}$ waveform linearly ramps the laser detuning by an additional 15 MHz over 12 ms to implement polarization-gradient cooling (PGC). After this ramp, all MOT and repump light is switched off at $t=250$ ms, initiating a time-of-flight measurement with delays ranging from 0 to 6 ms. A 0.5 ms illumination pulse, applied at $\Delta_{\mathrm{MOT}}$ (and optionally at $\Delta=0$ in future runs), captures the atomic fluorescence image. At $t=290$ ms, the system re-engages Rb referencing by first returning the PZT to the lock point and then re-enabling the SAS feedback.

Atomic cloud temperatures are determined from time-of-flight (TOF) measurements \cite{lett_observation_1988} (see Supplemental Materials section S2). Four test configurations are reported in Fig. \ref{fig:results_temperature}A. Three of the test configurations employ the dual-stage laser lock. These tests include: (I) neither frequency nor laser intensity PGC ramps (no intentional sub-Doppler cooling), (II) only a frequency PGC ramp (via PZT), and (III) both a frequency (via PZT) and laser intensity (via SOA) ramp. The fourth (IV) test configuration uses an AOM for a frequency ramp and the SOA for an intensity ramp. Figure \ref{fig:sup_schematics_during_TOF} illustrates the four cooling configurations and Figure \ref{fig:sup_timing_diagram} provides details of the experimental timing for configuration III. More discussion on the temperature data collection can be found in Supplementary Materials section S2.

The coldest measured temperature reaches $16 ~\mu \mathrm{K}$ via configuration (III). In principle, the AOM-based configuration (IV) should enable similarly low temperatures; however, optimization is more challenging because the laser coupling efficiency becomes detuning-dependent. As a result, the minimum temperature achieved using the AOM configuration was $28~\mu\mathrm{K}$.
    
\subsubsection*{Resonator agility measurements}

We characterize the frequency-agility of the PDH-locked PZT–SiN resonator by monitoring how the laser frequency tracks cavity tuning within the PDH feedback bandwidth. Figure \ref{fig:results_beatnote_control_monitor}A shows the measurement schematic. A 780-nm DBR cooling laser, PDH-locked to the PZT resonator, is heterodyned with a second 780-nm reference DBR laser stabilized to an independent Rb $\mathrm{D}_2$-line saturation-absorption spectroscopy (SAS) system. Using a fast photodiode, an electronic spectrum analyzer (ESA), and a frequency counter, we record the beat-note between the two lasers. Simultaneously, a camera monitors the fluorescence from the $^{87}$Rb MOT.

Figure \ref{fig:results_beatnote_control_monitor}B shows the laser beat-note recorded over 20 minutes of continuous operation. While we periodically referenced the PZT cavity to the Rb SAS (top), the beat-note remained stable near 78 MHz during the Rb lock and 185 MHz during MOT formation. We observed a slow drift in MOT fluorescence over long continuous operation (see extended atom number stability data in Figure \ref{fig:sup_hysteresis_and_atom_number}A). This single-MHz-scale, long-term drift is consistent with slow actuator dynamics (PZT creep and hysteresis) and/or residual thermal drift in the photonic package, which can shift the effective laser detuning from the optimum $\Delta_{\mathrm{MOT}}$ (see Supplemental Materials section S2). Importantly, the periodic re-referencing to the Rb SAS maintains repeatable frequency states and prevents runaway drift. In future operation, we aim to further suppress the residual fluorescence drift by adding a slow trim loop using MOT fluorescence as an error signal. Without Rb referencing (bottom), frequency drift accumulates rapidly, leading to the complete loss of the MOT signal within the first minute of operation. This confirms that the hybrid PDH with SAS scheme provides both short-term stability from the cavity and long-term drift suppression from the atomic reference. In separate measurement runs without simultaneous beat-note monitoring, the MOT atom number remained stable for at least 30 minutes, indicating that PZT hysteresis was negligible under these operating conditions (\ref{fig:sup_hysteresis_and_atom_number}B).

The zoomed-in section of three sequential cycles, shown in Fig.~\ref{fig:results_beatnote_control_monitor}C, highlights the repeatable frequency trajectory of each cycle. Figure \ref{fig:results_beatnote_control_monitor}D overlays 100 consecutive cycles, showing consistent operation over the entire PZT drive range. Each sequence begins with a 25 ms Rb referencing interval followed by the MOT stage and concludes with a 12 ms triangular frequency ramp corresponding to the polarization-gradient cooling (PGC) sequence.

 Finally, Fig.~\ref{fig:results_beatnote_control_monitor}E,F illustrate the flexibility of the system: an arbitrary waveform applied to the PZT produces a programmed laser-frequency trajectory recreating a two-dimensional pattern. This highlights the ability of the integrated PZT–SiN resonator to provide deterministic, high-speed, and repeatable frequency control over hundreds of MHz. Together, these results confirm that the resonator enables robust, long-term frequency holdover when referenced to Rb spectroscopy and precise, agile tuning on millisecond timescales—sufficient for implementing sub-Doppler cooling and dynamic frequency control in cold-atom experiments.

\subsection*{Discussion}

We have demonstrated an integrated 780 nm PZT-on-SiN resonator that enables programmable, precise, and agile laser frequency control for cold-atom applications. By locking a 780-nm DBR laser to the programmable high-Q (2.8 million intrinsic) resonator and driving the PZT actuator with sub-doppler cooling voltage sequences, we cool rubidium atoms to 16 $\mathrm{\mu K}$ in a 3D MOT using rapid 15-MHz linear detuning over 12 ms for polarization-gradient cooling (PGC). In this demonstration, the PZT-based agile cavity fully replaces the role of the acousto-optic modulator (AOM) in the sub-Doppler cooled 3D-MOT. This approach maintains stable laser lock and achieves atomic temperatures far below the 146 $\mathrm{\mu K}$ Doppler limit. These results demonstrate that wideband stress-optic tuning can provide both long-term stability and fast, programmable frequency agility within a single integrated photonic component. This approach reduces power consumption and mitigates issues with conventional RF-based approaches such as acousto-optic or electro-optic modulation that generate unwanted optical sidebands and can parasitically excite nearby atomic transitions. By programming the optical reference resonator itself, the laser remains single-frequency and inherits a deterministic frequency trajectory from the high-Q resonator. Furthermore, this scheme is compatible with frequency noise reduction and linewidth narrowing due to the low thermo-refractive noise (TRN), high Q, and high extinction ratio (ER) of the resonator~\cite{isichenko_sub-hz_2024, isichenko_multi-laser_2025}. This sideband-free frequency synthesis eliminates spurious optical tones during the cooling sequence, simplifying system architecture and reducing undesired atomic excitation.

The SiN waveguide layer here matches the layer used in photonic-integrated beam-delivery systems for magneto-optical traps \cite{isichenko_photonic_2023}, underscoring the compatibility of this approach with large-area emitter arrays and integrated cold-atom platforms. Beyond frequency tuning, the same PZT actuation mechanism can be engineered for direct amplitude modulation and optical shuttering, either through interferometric or resonant configurations. Previous demonstrations of multi-ring and Mach–Zehnder interferometer (MZI) modulators in SiN and AlN platforms \cite{huffman_integrated_2018, xie_aluminum_2025} indicate that similar stress-optic actuation could enable fast amplitude modulation and switching. In parallel, semiconductor optical amplifiers (SOAs) have recently been demonstrated as integrated amplitude modulators and shutters \cite{kittlaus_semiconductor_2025}, providing a complementary path toward all-on-chip control of laser power and timing.

Once disciplined to the atomic reference, the resonator maintains frequency stability across the millisecond-scale cooling window without re-referencing, enabling reproducible 100 MHz frequency ramps over hundreds of cycles. This highlights the cavity’s intrinsic ability to serve as a short-term holdover cavity, which itself arises from the excellent short-term coherence provided by the high-Q SiN cavity and its low-power PZT actuation. This functionality eliminates the need for bulk optic reference cavities and further enables integration, compactness, and low-power operation.

Looking forward, this stress-optic tuning approach is compatible with high-bandwidth self-injection-locked (SIL) architectures \cite{corato-zanarella_widely_2022, isichenko_sub-hz_2024}, where direct feedback between the PZT-tuned resonator and the laser can further enhance frequency stability and tuning range. The PZT-actuated resonator thus represents one element of a broader set of integrated photonic components required for atomic and quantum systems. When combined with narrow-linewidth external-cavity lasers at 852 nm \cite{nejadriahi_sub-100_2024}, self-injection-locked lasers at 780 nm \cite{isichenko_sub-hz_2024}, and integrated vapor-cell spectroscopy references, this platform can enable fully photonic implementations of cold-atom sensors and clocks. Moreover, we may perform the rubidium referencing using either a vapor-cell SAS or the MOT itself, as in self-locked MOT architectures \cite{fletcher_self-locked_2002}, offering a compact route to autonomous frequency stabilization.

\begin{figure} 
	\centering
	\includegraphics[width=0.95\textwidth]{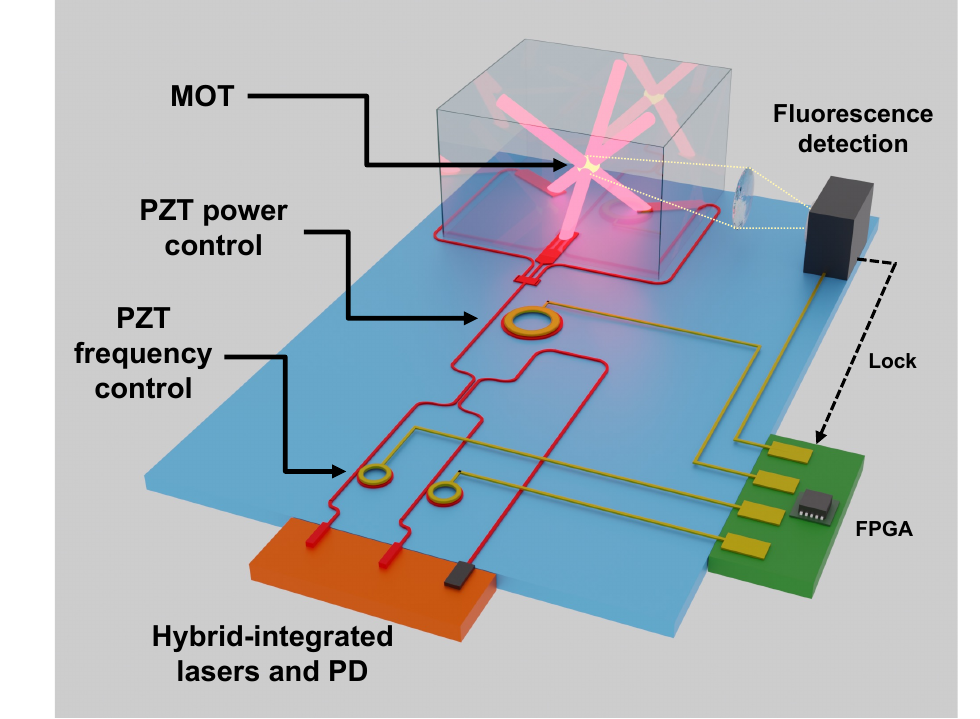} % for an image file named 
	\caption{\textbf{Example integrated cold atom system enabled by the 780 PZT resonators.} 
    A pair of self-injection locked lasers consisting of hybrid-integrated edge-emitting lasers and PZT-on-SiN resonators form the frequency-agile cooling and repump lasers for a MOT. The frequency spacing between the lasers is controlled with an offset beat-note lock realized with a directional coupler splitter and a photodetector. The power shuttering and ramping is controlled with a critically-coupled PZT resonator. The absolute frequency of the cooling laser is set based on feedback from the MOT fluorescence.} 
	\label{fig:vision}
\end{figure}

Figure~\ref{fig:vision} presents our vision for a fully integrated cold-atom control system enabled by the fast-tunable PZT–SiN resonators we demonstrate here. In this architecture, a pair of self-injection-locked 780-nm lasers provides the cooling and repump light. Independent PZT actuators stabilize and tune each laser. The frequency spacing between the lasers is set by an on-chip offset beat-note lock using an integrated directional coupler and photodetector. We aim to implement power ramping and shuttering through critically coupled resonators driven by PZT amplitude control, while fluorescence feedback from the magneto-optical trap (MOT) provides absolute frequency referencing. Combined with integrated electronics for data processing and feedback (e.g., FPGA-based control), this system concept represents a path toward a chip-scale cold-atom instrument where laser frequency, power, and atomic readout are all governed by the same photonic platform.

In summary, the programmable, agile, PZT-on-SiN resonator combines low optical loss, high tuning efficiency, wideband response, low power consumption, and the ability to operate as a holdover cavity and perform repeatable, rapid, and programmable frequency tuning for atomic quantum experiments. The successful demonstration of sub-Doppler cooling establishes a new paradigm for integrated frequency control in atomic physics. By integrating laser stabilization, agile modulation, and atomic referencing on a single photonic platform, this work paves the way toward compact, power-efficient, and scalable cold-atom instruments for quantum sensing and precision metrology.

\subsection*{Materials and Methods}

\subsubsection*{Device fabrication and packaging}

We fabricate the PZT-on-SiN resonator device waveguides on a 15-$\mu$m thick lower cladding layer consisting of thermal silica dioxide on a 1-mm thick, 100-mm diameter silicon substrate. Next, a 120-nm thick stoichiometric silicon nitride layer is deposited by low-pressure chemical vapor deposition (LPCVD) directly on top of the thermal oxide layer. The waveguides in the silicon-nitride layer are patterned using a photoresist mask with a 248-nm DUV stepper lithography tool and an inductively coupled plasma etch with CF4/CHF3/O2 gas. Following the pattern etch, we deposit a 4-$\mu$m thick layer of SiO$_2$ upper cladding by plasma enhanced chemical vapor deposition (PECVD) with tetraethylorthosilicate (TEOS) used as a precursor. The wafer is annealed at 1050$^{\circ}$C for 7 hours and at  1150$^{\circ}$C for 2 hours. The PZT actuator stack consists of a sputtered 40-nm thick TiO$_2$ adhesion layer, a sputtered 150-nm thick Pt bottom electrode, and a 1-$\mu$m thick layer of PZT (52/48 Zr/Ti ratio) deposited via chemical solution deposition (CSD). The stack is capped with a sputtered 100-nm thick Pt top electrode. The PZT and Pt electrodes are patterned by argon ion milling. With the actuator patterned, electrical traces are evaporated and patterned through lift-off and consist of a Cr/Pt/Au stack with thicknesses of 20 nm, 20 nm, and 730 nm respectively. To reduce resistivity and minimize gold coverage on the electrodes, a 10-$\mu$m thick copper layer is electroplated using photoresist molds and a sputtered copper seed layer. The photoresist molds and copper seed layer are removed with solvents to release the device.

The wafer is diced to isolate several ring resonator devices from each wafer reticle. The die is bonded with thermally conductive epoxy on top of a thermoelectric cooler (TEC) mounted inside an aluminum enclosure. A thermistor is placed on the TEC for temperature stabilization. A pair of polarization-maintaining single-fiber v-grooves is aligned to the resonator bus and bonded in place with UV-cured epoxy. Electrical contacts to the PZT actuator are made with gold wire-bonds to a circuit board with an electrical feed-through to outside the enclosure. During MOT operation and beat-note measurements, a Vescent$^{\rm{TM}}$ SLICE temperature controller maintains a resonator stability of $\le 0.1 ~\mathrm{mK}$.

%%%%%%%%%%%%%%%% MAIN TEXT FIGURES %%%%%%%%%%%%%%%

% TO ADD A FIGURE:
% use this google drive format
% https://drive.google.com/uc?export=download&id=FILE_ID
% and plug in the FILE_ID from the copy-link feature. file must be set to "anyone with the link"
% notes: https://docs.overleaf.com/integrations-and-add-ons/google-drive

\begin{figure} 
	\centering
	\includegraphics[width=0.95\textwidth]{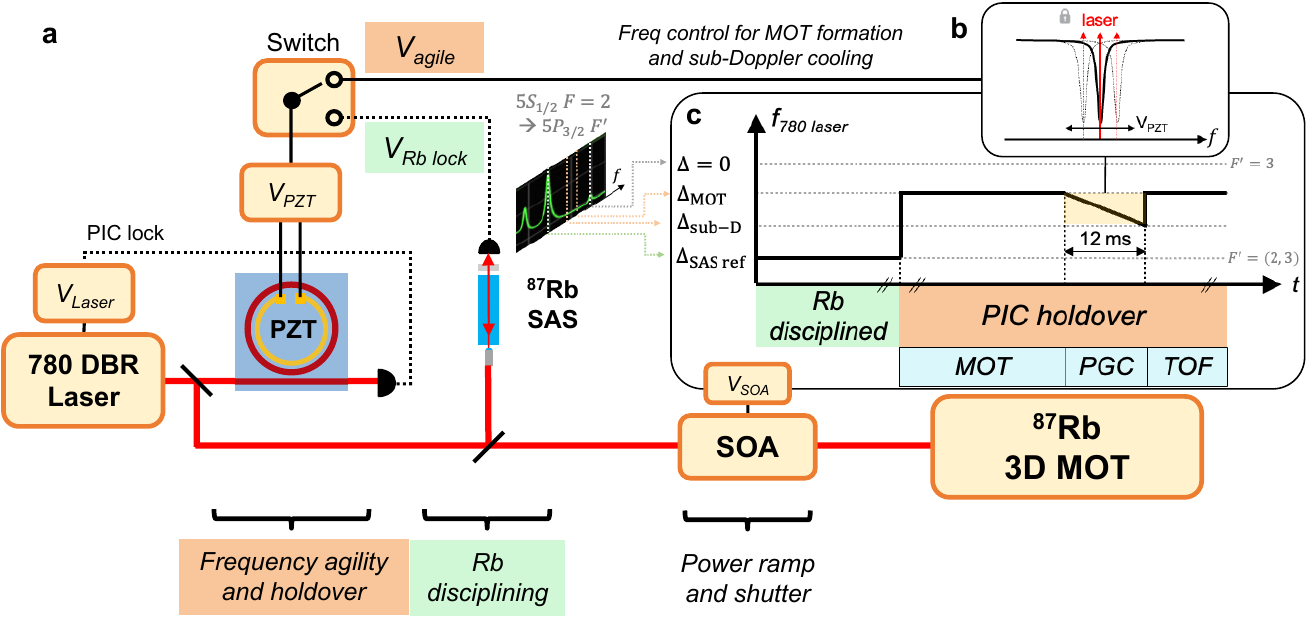}
	\caption{\textbf{Cold atom experiment diagram.}
		  \textbf{a)} The 780-nm DBR cooling laser is stabilized to the PZT PIC resonator cavity for reduced laser drift (i.e. holdover) and locked laser frequency control applied with $V_{\rm PZT}$. The PZT control switches between fast frequency control ($V_{\rm agile}$) used to control the cavity for 1) Rb disciplining, by locking to the Rb saturation absorption spectroscopy (SAS) and 2) MOT formation and for sub-Doppler cooling. The cooling laser light is amplified with a semiconductor optical amplifier (SOA) which is voltage-controlled ($V_{\rm SOA}$) for power ramping and shuttering, used for sub-Doppler cooling temperature measurements.  \textbf{b)} The $V_{\rm agile}$ frequency control tunes the cooling laser frequency enabling arbitrary laser frequency control within the tuning range of the PZT actuator and at speeds within the locking range of the laser lock. \textbf{c)} Timing diagram for one cold atom experiment cycle consisting of the Rb disciplining of the PZT PIC cavity and tuning the PIC to set the cooling laser frequency with respect to Rb hyperfine transitions. The Rb disciplining is done at the strong $F'=(2,3)$ cross-over transition for 25 ms (see Fig. ~\ref{fig:sup_timing_diagram}). The PZT is jumped by a set voltage to set the cooling laser at $\Delta_{\rm MOT}$ detuning from the cooling ($F'=3$) transition. After full MOT loading, the fast frequency control signal is applied to the PZT to rapidly ramp the laser by 15 MHz in 12 ms for polarization gradient cooling (PGC) to achieve sub-Doppler cooling, measured during the time-of-flight (TOF) stage. 
    }
	\label{fig:exp_diagram}
\end{figure}

\begin{figure} 
	\centering
	\includegraphics[width=0.95\textwidth]{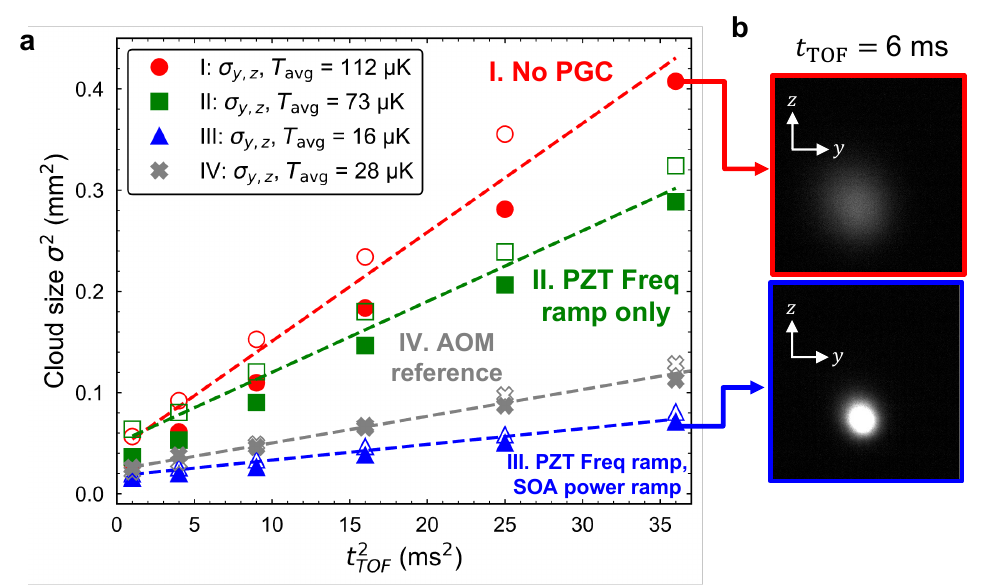} 
	\caption{\textbf{MOT temperature measurements for different cooling laser control configurations.}
		  \textbf{a)} Time-of-flight (TOF) atom cloud temperature measurements, with the squared cloud radius along the $z$ (open markers) and $y$ (filled markers) dimensions for different squared TOF times $t_{\rm TOF}^2$. The linear fits are for $\sigma^2_i = \sigma^2_{i,0} + \frac{k_B T_i}{m} t_{\rm TOF}^2$, where $\sigma_i$ is the Gaussian standard deviation of the cloud along an axis ($y,z$), $\sigma_{i,0}$ is the initial width of the cloud, $k_B$ is the Boltzmann constant, $m$ is the mass of a single $^{87}$Rb atom, and $T_i$ is the temperature along an axis. The schematic for the experiment configurations (I-IV) is shown in Figure \ref{fig:sup_schematics_during_TOF}. \textbf{c)} Fluorescence images of the MOT cloud after free-expansion for $t_{\rm TOF}$ = 6 ms. The color scale is identical for both images. 
}
	\label{fig:results_temperature}
\end{figure}

\begin{figure} 
	\centering
	\includegraphics[width=1.0\textwidth]{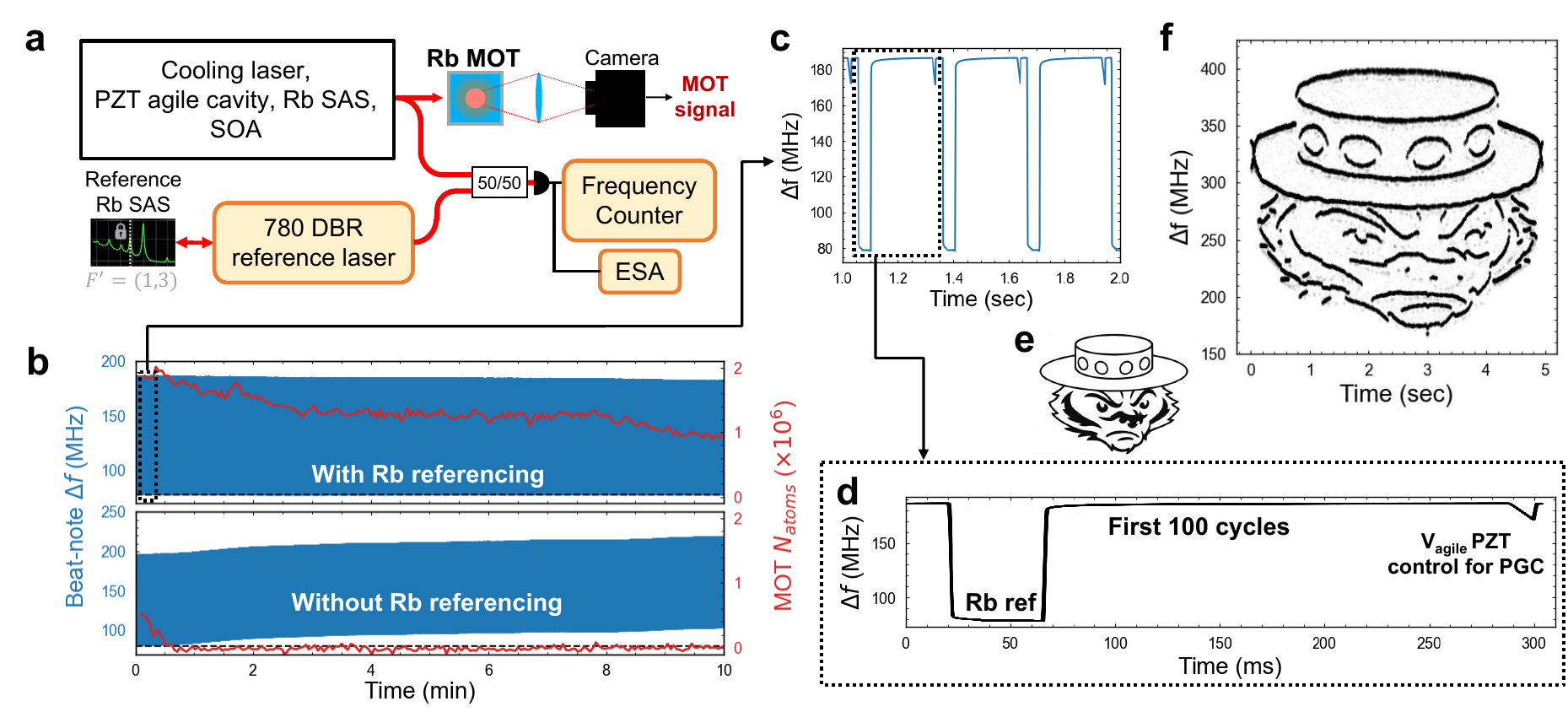} 
	\caption{\textbf{Cooling laser frequency agility measurements.}
		  \textbf{a)} Schematic for monitoring the PZT-controlled cooling laser frequency during the MOT and sub-Doppler cooling. A reference 780-nm DBR laser is locked to another Rb spectroscopy system and the heterodyne beat-note between the reference and the cooling lasers is recorded on a photodiode and a frequency counter. During the sequence the beat-note is monitored on an electronic spectrum analyzer (ESA) and the MOT fluorescence signal is extracted from a camera. \textbf{b)} Laser beat-note recorded on a frequency counter during 20 minutes of continuous operation. When the PZT cavity is re-referenced to the Rb SAS (with Rb reference, top), the beat-note remains near 78 MHz (185 MHz) for the referencing (MOT formation) parts of the cycle. We record the MOT atom number the entire time. Without the Rb referencing (bottom) the drift results in the loss of trapped atom within the first minute of operation. \textbf{c)} Zoom-in of the beat-note in (b, top) for six cycles. \textbf{(d)} One period from of (c) overlaid with 100 consecutive cycles, showing the stability of the PZT response during this period. The first jump has a ~25-ms Rb referencing time and the triangular ramp at 290 ms is the PGC ramp. \textbf{(e)} Pre-programmed reference image illustrating the intended laser beat-note trajectory shown in \textbf{(f)}.} 
	\label{fig:results_beatnote_control_monitor}
\end{figure}

%%%%%%%%%%%%%%%% REFERENCES %%%%%%%%%%%%%%%

\clearpage % Clear all remaining figures and tables then start a new page

\bibliography{references} % for a file named references.bib

@article{stray_quantum_2022,
	title = {Quantum sensing for gravity cartography},
	volume = {602},
	rights = {2022 The Author(s)},
	issn = {1476-4687},
	url = {https://www.nature.com/articles/s41586-021-04315-3},
	doi = {10.1038/s41586-021-04315-3},
	pages = {590--594},
	number = {7898},
	 journal = {Nature},
	author = {Stray, Ben and Lamb, Andrew and Kaushik, Aisha and Vovrosh, Jamie and Rodgers, Anthony and Winch, Jonathan and Hayati, Farzad and Boddice, Daniel and Stabrawa, Artur and Niggebaum, Alexander and Langlois, Mehdi and Lien, Yu-Hung and Lellouch, Samuel and Roshanmanesh, Sanaz and Ridley, Kevin and de Villiers, Geoffrey and Brown, Gareth and Cross, Trevor and Tuckwell, George and Faramarzi, Asaad and Metje, Nicole and Bongs, Kai and Holynski, Michael},
	urldate = {2022-02-23},
	date = {2022-02},
    year = {2022},
	langid = {english},
}

@article{lee_compact_2022,
	title = {A compact cold-atom interferometer with a high data-rate grating magneto-optical trap and a photonic-integrated-circuit-compatible laser system},
	volume = {13},
	rights = {2022 This is a U.S. Government work and not under copyright protection in the {US}; foreign copyright protection may apply},
	issn = {2041-1723},
	url = {https://www.nature.com/articles/s41467-022-31410-4},
	doi = {10.1038/s41467-022-31410-4},
	pages = {5131},
	number = {1},
	journal = {Nature Communications},
	shortjournal = {Nat Commun},
	author = {Lee, Jongmin and Ding, Roger and Christensen, Justin and Rosenthal, Randy R. and Ison, Aaron and Gillund, Daniel P. and Bossert, David and Fuerschbach, Kyle H. and Kindel, William and Finnegan, Patrick S. and Wendt, Joel R. and Gehl, Michael and Kodigala, Ashok and {McGuinness}, Hayden and Walker, Charles A. and Kemme, Shanalyn A. and Lentine, Anthony and Biedermann, Grant and Schwindt, Peter D. D.},
	urldate = {2022-09-01},
	date = {2022-09-01},
    year = {2022},
	langid = {english},
}

@article{ludlow_optical_2015,
	title = {Optical atomic clocks},
	volume = {87},
	url = {https://link.aps.org/doi/10.1103/RevModPhys.87.637},
	doi = {10.1103/RevModPhys.87.637},
	pages = {637--701},
	number = {2},
	journal = {Reviews of Modern Physics},
	shortjournal = {Rev. Mod. Phys.},
	author = {Ludlow, Andrew D. and Boyd, Martin M. and Ye, Jun and Peik, E. and Schmidt, P. O.},
	urldate = {2022-08-02},
    year = {2015},
	date = {2015-06-26},
}

@article{graham_multi-qubit_2022,
	title = {Multi-qubit entanglement and algorithms on a neutral-atom quantum computer},
	volume = {604},
	rights = {2022 The Author(s), under exclusive licence to Springer Nature Limited},
	issn = {1476-4687},
	url = {https://www.nature.com/articles/s41586-022-04603-6},
	doi = {10.1038/s41586-022-04603-6},
	pages = {457--462},
	number = {7906},
	 journal = {Nature},
	author = {Graham, T. M. and Song, Y. and Scott, J. and Poole, C. and Phuttitarn, L. and Jooya, K. and Eichler, P. and Jiang, X. and Marra, A. and Grinkemeyer, B. and Kwon, M. and Ebert, M. and Cherek, J. and Lichtman, M. T. and Gillette, M. and Gilbert, J. and Bowman, D. and Ballance, T. and Campbell, C. and Dahl, E. D. and Crawford, O. and Blunt, N. S. and Rogers, B. and Noel, T. and Saffman, M.},
	urldate = {2022-08-02},
	date = {2022-04},
    year = {2022},
	langid = {english},
}

@article{chiow_compact_2018,
	title = {Compact atom interferometer using single laser},
	volume = {124},
	issn = {1432-0649},
	url = {https://doi.org/10.1007/s00340-018-6965-2},
	doi = {10.1007/s00340-018-6965-2},
	pages = {96},
	number = {6},
	 journal = {Applied Physics B},
	shortjournal = {Appl. Phys. B},
	author = {Chiow, Sheng-wey and Yu, Nan},
	urldate = {2021-06-28},
	date = {2018-05-08},
    year = {2018},
	langid = {english},
}

@article{kodigala_high-performance_2024,
	title = {High-performance silicon photonic single-sideband modulators for cold-atom interferometry},
	volume = {10},
	url = {https://www.science.org/doi/10.1126/sciadv.ade4454},
	doi = {10.1126/sciadv.ade4454},
	pages = {eade4454},
	number = {28},
	 journal = {Science Advances},
	author = {Kodigala, Ashok and Gehl, Michael and Hoth, Gregory W. and Lee, Jongmin and {DeRose}, Christopher T. and Pomerene, Andrew and Dallo, Christina and Trotter, Douglas and Starbuck, Andrew L. and Biedermann, Grant and Schwindt, Peter D. D. and Lentine, Anthony L.},
	urldate = {2024-08-14},
	date = {2024-07-10},
    year = {2024},
}

@article{puckett_422_2021,
	title = {422 Million intrinsic quality factor planar integrated all-waveguide resonator with sub-{MHz} linewidth},
	volume = {12},
	rights = {2021 The Author(s)},
	issn = {2041-1723},
	url = {https://www.nature.com/articles/s41467-021-21205-4},
	doi = {10.1038/s41467-021-21205-4},
	pages = {934},
	number = {1},
	 journal = {Nature Communications},
	shortjournal = {Nat Commun},
	author = {Puckett, Matthew W. and Liu, Kaikai and Chauhan, Nitesh and Zhao, Qiancheng and Jin, Naijun and Cheng, Haotian and Wu, Jianfeng and Behunin, Ryan O. and Rakich, Peter T. and Nelson, Karl D. and Blumenthal, Daniel J.},
	urldate = {2021-06-29},
	date = {2021-02-10},
    year = {2021},
	langid = {english},
}

@article{blumenthal_silicon_2018,
	title = {Silicon Nitride in Silicon Photonics},
	volume = {106},
	rights = {https://ieeexplore.ieee.org/Xplorehelp/downloads/license-information/{OAPA}.html},
	issn = {0018-9219, 1558-2256},
	url = {https://ieeexplore.ieee.org/document/8472140/},
	doi = {10.1109/JPROC.2018.2861576},
	pages = {2209--2231},
	number = {12},
	 journal = {Proceedings of the {IEEE}},
	shortjournal = {Proc. {IEEE}},
	author = {Blumenthal, Daniel J. and Heideman, Rene and Geuzebroek, Douwe and Leinse, Arne and Roeloffzen, Chris},
	urldate = {2024-04-30},
	date = {2018-12},
    year = {2018},
	langid = {english},
}

@article{alexander_nanophotonic_2018,
	title = {Nanophotonic Pockels modulators on a silicon nitride platform},
	volume = {9},
	rights = {2018 The Author(s)},
	issn = {2041-1723},
	url = {https://www.nature.com/articles/s41467-018-05846-6.},
	doi = {10.1038/s41467-018-05846-6},
	pages = {3444},
	number = {1},
	 journal = {Nature Communications},
	shortjournal = {Nat Commun},
	author = {Alexander, Koen and George, John P. and Verbist, Jochem and Neyts, Kristiaan and Kuyken, Bart and Van Thourhout, Dries and Beeckman, Jeroen},
	urldate = {2023-08-19},
	date = {2018-08-27},
    year = {2018},
	langid = {english},
}

@article{wang_silicon_2022,
	title = {Silicon nitride stress-optic microresonator modulator for optical control applications},
	volume = {30},
	rights = {\&\#169; 2022 Optica Publishing Group},
	issn = {1094-4087},
	url = {https://opg.optica.org/oe/abstract.cfm?uri=oe-30-18-31816},
	doi = {10.1364/OE.467721},
	pages = {31816--31827},
	number = {18},
	 journal = {Optics Express},
	shortjournal = {Opt. Express, {OE}},
	author = {Wang, Jiawei and Liu, Kaikai and Harrington, Mark W. and Rudy, Ryan Q. and Blumenthal, Daniel J.},
	urldate = {2022-08-18},
	date = {2022-08-29},
    year = {2022},
}

@article{stanfield_cmos-compatible_2019,
	title = {{CMOS}-compatible, piezo-optomechanically tunable photonics for visible wavelengths and cryogenic temperatures},
	volume = {27},
	rights = {\&\#169; 2019 Optical Society of America},
	issn = {1094-4087},
	url = {https://opg.optica.org/oe/abstract.cfm?uri=oe-27-20-28588},
	doi = {10.1364/OE.27.028588},
	pages = {28588--28605},
	number = {20},
	 journal = {Optics Express},
	shortjournal = {Opt. Express, {OE}},
	author = {Stanfield, P. R. and Leenheer, A. J. and Michael, C. P. and Sims, R. and Eichenfield, M.},
	urldate = {2024-04-30},
	date = {2019-09-30},
    year = {2019},
}

@article{dalibard_laser_1989,
	title = {Laser cooling below the Doppler limit by polarization gradients: simple theoretical models},
	volume = {6},
	rights = {\&\#169; 1989 Optical Society of America},
	issn = {1520-8540},
	url = {https://opg.optica.org/josab/abstract.cfm?uri=josab-6-11-2023},
	doi = {10.1364/JOSAB.6.002023},
	shorttitle = {Laser cooling below the Doppler limit by polarization gradients},
	pages = {2023--2045},
	number = {11},
	 journal = {{JOSA} B},
	shortjournal = {J. Opt. Soc. Am. B, {JOSAB}},
	author = {Dalibard, J. and Cohen-Tannoudji, C.},
	urldate = {2023-11-05},
	date = {1989-11-01},
    year = {1989},
}

@misc{isichenko_multi-laser_2025,
	title = {Multi-laser stabilization with an atomic-disciplined photonic integrated resonator},
	url = {http://arxiv.org/abs/2509.09124},
	doi = {10.48550/arXiv.2509.09124},
	number = {{arXiv}:2509.09124},
	publisher = {{arXiv}},
	author = {Isichenko, Andrei and Hunter, Andrew S. and Chauhan, Nitesh and Dickson, John R. and Nunley, T. Nathan and Bingaman, Josiah R. and Heim, David A. S. and Harrington, Mark W. and Liu, Kaikai and Kunz, Paul D. and Blumenthal, Daniel J.},
	urldate = {2025-09-24},
	date = {2025-09-17},
    year = {2025},
	eprinttype = {arxiv},
	eprint = {2509.09124 [physics]},
}

@article{corato-zanarella_widely_2022,
	title = {Widely tunable and narrow-linewidth chip-scale lasers from near-ultraviolet to near-infrared wavelengths},
	rights = {2022 The Author(s), under exclusive licence to Springer Nature Limited},
	issn = {1749-4893},
	url = {https://www.nature.com/articles/s41566-022-01120-w},
	doi = {10.1038/s41566-022-01120-w},
	pages = {1--8},
	 journal = {Nature Photonics},
	shortjournal = {Nat. Photon.},
	author = {Corato-Zanarella, Mateus and Gil-Molina, Andres and Ji, Xingchen and Shin, Min Chul and Mohanty, Aseema and Lipson, Michal},
	urldate = {2022-12-25},
	date = {2022-12-23},
    year = {2022},
	langid = {english},
}

@article{isichenko_photonic_2023,
	title = {Photonic integrated beam delivery for a rubidium 3D magneto-optical trap},
	volume = {14},
	rights = {2023 The Author(s)},
	issn = {2041-1723},
	url = {https://www.nature.com/articles/s41467-023-38818-6},
	doi = {10.1038/s41467-023-38818-6},
	pages = {3080},
	number = {1},
	 journal = {Nature Communications},
	shortjournal = {Nat Commun},
	author = {Isichenko, Andrei and Chauhan, Nitesh and Bose, Debapam and Wang, Jiawei and Kunz, Paul D. and Blumenthal, Daniel J.},
	urldate = {2023-05-29},
	date = {2023-05-29},
    year = {2023},
	langid = {english},
}

@article{isichenko_sub-hz_2024,
	title = {Sub-Hz fundamental, sub-{kHz} integral linewidth self-injection locked 780 nm hybrid integrated laser},
	volume = {14},
	rights = {2024 The Author(s)},
	issn = {2045-2322},
	url = {https://www.nature.com/articles/s41598-024-76699-x},
	doi = {10.1038/s41598-024-76699-x},
	pages = {27015},
	number = {1},
	 journal = {Scientific Reports},
	shortjournal = {Sci Rep},
	author = {Isichenko, Andrei and Hunter, Andrew S. and Bose, Debapam and Chauhan, Nitesh and Song, Meiting and Liu, Kaikai and Harrington, Mark W. and Blumenthal, Daniel J.},
	urldate = {2024-11-18},
	date = {2024-11-18},
    year = {2024},
	langid = {english},
}

@article{wang_integrated_2023,
	title = {Integrated programmable strongly coupled three-ring resonator photonic molecule with ultralow-power piezoelectric control},
	volume = {48},
	rights = {© 2023 Optica Publishing Group},
	issn = {1539-4794},
	url = {https://opg.optica.org/ol/abstract.cfm?uri=ol-48-9-2373},
	doi = {10.1364/OL.482567},
	pages = {2373--2376},
	number = {9},
	 journal = {Optics Letters},
	shortjournal = {Opt. Lett., {OL}},
	author = {Wang, Jiawei and Liu, Kaikai and Isichenko, Andrei and Rudy, Ryan Q. and Blumenthal, Daniel J.},
	urldate = {2023-04-26},
	date = {2023-05-01},
    year = {2023},
}

@article{nejadriahi_sub-100_2024,
	title = {Sub-100 Hz intrinsic linewidth 852 nm silicon nitride external cavity laser},
	volume = {49},
	rights = {© 2024 Optica Publishing Group},
	issn = {1539-4794},
	url = {https://opg.optica.org/ol/abstract.cfm?uri=ol-49-24-7254},
	doi = {10.1364/OL.543307},
	pages = {7254--7257},
	number = {24},
	 journal = {Optics Letters},
	shortjournal = {Opt. Lett., {OL}},
	author = {Nejadriahi, Hani and Kittlaus, Eric and Bose, Debapam and Chauhan, Nitesh and Wang, Jiawei and Fradet, Mathieu and Bagheri, Mahmood and Isichenko, Andrei and Heim, David and Forouhar, Siamak and Blumenthal, Daniel J.},
	urldate = {2025-05-30},
	date = {2024-12-15},
    year = {2024},
}

@article{fletcher_self-locked_2002,
	title = {A self-locked magneto-optic trap},
	volume = {212},
	issn = {0030-4018},
	url = {https://www.sciencedirect.com/science/article/pii/S0030401802019521},
	doi = {10.1016/S0030-4018(02)01952-1},
	pages = {85--88},
	number = {1},
	 journal = {Optics Communications},
	shortjournal = {Optics Communications},
	author = {Fletcher, C. S and Lye, J. E and Robins, N. P and Close, J. D},
	urldate = {2024-08-30},
	date = {2002-10-15},
    year = {2002},
}

@article{rosi_-enhanced_2018,
	title = {{$\Lambda$}-enhanced grey molasses on the D2 transition of Rubidium-87 atoms},
	volume = {8},
	rights = {2018 The Author(s)},
	issn = {2045-2322},
	url = {https://www.nature.com/articles/s41598-018-19814-z},
	doi = {10.1038/s41598-018-19814-z},
	pages = {1301},
	number = {1},
	 journal = {Scientific Reports},
	shortjournal = {Sci Rep},
	author = {Rosi, Sara and Burchianti, Alessia and Conclave, Stefano and Naik, Devang S. and Roati, Giacomo and Fort, Chiara and Minardi, Francesco},
	urldate = {2023-09-11},
	date = {2018-01-22},
    year = {2018},
	langid = {english},
}

@article{jin_piezoelectrically_2018,
	title = {Piezoelectrically tuned silicon nitride ring resonator},
	volume = {26},
	rights = {\&\#169; 2018 Optical Society of America},
	issn = {1094-4087},
	url = {https://opg.optica.org/oe/abstract.cfm?uri=oe-26-3-3174},
	doi = {10.1364/OE.26.003174},
	pages = {3174--3187},
	number = {3},
	 journal = {Optics Express},
	shortjournal = {Opt. Express, {OE}},
	author = {Jin, Warren and Polcawich, Ronald G. and Morton, Paul A. and Bowers, John E.},
	urldate = {2024-10-20},
	date = {2018-02-05},
    year = {2018},
}

@article{kittlaus_semiconductor_2025,
	title = {Semiconductor optical amplifier-based laser system for cold-atom sensors},
	volume = {12},
	rights = {2025 This is a U.S. Government work and not under copyright protection in the {US}; foreign copyright protection may apply},
	issn = {2196-0763},
	url = {https://epjquantumtechnology.springeropen.com/articles/10.1140/epjqt/s40507-025-00348-z},
	doi = {10.1140/epjqt/s40507-025-00348-z},
	pages = {1--19},
	number = {1},
	 journal = {{EPJ} Quantum Technology},
	shortjournal = {{EPJ} Quantum Technol.},
	author = {Kittlaus, Eric and Hunacek, Jonathon and Bagheri, Mahmood and Nejadriahi, Hani and Langlois, Mehdi and Chiow, Sheng-wey and Yu, Nan and Forouhar, Siamak},
	urldate = {2025-10-16},
	date = {2025-12},
    year = {2025},
	langid = {english},
}

@article{huffman_integrated_2018,
	title = {Integrated Resonators in an Ultralow Loss Si3N4/{SiO}2 Platform for Multifunction Applications},
	volume = {24},
	issn = {1558-4542},
	doi = {10.1109/JSTQE.2018.2818459},
	pages = {1--9},
	number = {4},
	journal = {{IEEE} Journal of Selected Topics in Quantum Electronics},
	author = {Huffman, Taran Arthur and Brodnik, Grant M. and Pinho, Cátia and Gundavarapu, Sarat and Baney, Douglas and Blumenthal, Daniel J.},
	date = {2018-07},
    year = {2018},
}

@misc{siddharth_narrow-linewidth_2025,
	title = {Narrow-linewidth, piezoelectrically tunable photonic integrated blue laser},
	url = {http://arxiv.org/abs/2508.02568},
	doi = {10.48550/arXiv.2508.02568},
	number = {{arXiv}:2508.02568},
	publisher = {{arXiv}},
	author = {Siddharth, Anat and Gardner, Asger B. and Ji, Xinru and Hulyal, Shivaprasad U. and Reichler, Mikael S. and Attanasio, Alaina and Riemensberger, Johann and Bhave, Sunil A. and Volet, Nicolas and Bianconi, Simone and Kippenberg, Tobias J.},
	urldate = {2025-10-20},
	date = {2025-08-04},
    year = {2025},
	eprinttype = {arxiv},
	eprint = {2508.02568 [physics]},
}

@article{xie_aluminum_2025,
	title = {Aluminum nitride photonics for high-speed electro-optical tuning of self-injection-locked laser},
	volume = {50},
	issn = {0146-9592, 1539-4794},
	url = {https://opg.optica.org/abstract.cfm?URI=ol-50-18-5913},
	doi = {10.1364/OL.571314},
	pages = {5913},
	number = {18},
	 journal = {Optics Letters},
	shortjournal = {Opt. Lett.},
	author = {Xie, Hao and Wang, Yubo and Zhou, Yiyu and Holguín-Lerma, Jorge A. and Yang, Guangcanlan and Shi, Zhimin and Tang, Hong X.},
	urldate = {2025-09-17},
	date = {2025-09-15},
    year = {2025},
	langid = {english},
}

@inproceedings{montifiore_integrated_2025,
	title = {Integrated Low-Power Blue Light {PZT} Silicon Nitride Ring Modulator for Atomic and Quantum Applications},
	rights = {© 2025 The Author(s)},
	url = {https://opg.optica.org/abstract.cfm?uri=CLEO_SI-2025-SS160_5},
	doi = {10.1364/CLEO_SI.2025.SS160_5},
	eventtitle = {{CLEO}: Science and Innovations},
	pages = {SS160\_5},
	booktitle = {{CLEO} 2025, paper {SS}160\_5},
	publisher = {Optica Publishing Group},
	author = {Montifiore, Nick and Isichenko, Andrei and Wang, Jiawei and Chauhan, Nitesh and Harrington, Mark W. and Pushkarsky, Michael and Blumenthal, Daniel J.},
	urldate = {2025-10-24},
	date = {2025-05-04},
    year = {2025},
}

@article{liang_compact_2015,
	title = {Compact stabilized semiconductor laser for frequency metrology},
	volume = {54},
	issn = {0003-6935, 1539-4522},
	url = {https://opg.optica.org/abstract.cfm?URI=ao-54-11-3353},
	doi = {10.1364/AO.54.003353},
	pages = {3353},
	number = {11},
	 journal = {Applied Optics},
	shortjournal = {Appl. Opt.},
	author = {Liang, Wei and Ilchenko, Vladimir S. and Eliyahu, Danny and Dale, Elijah and Savchenkov, Anatoliy A. and Seidel, David and Matsko, Andrey B. and Maleki, Lute},
	urldate = {2022-05-10},
	date = {2015-04-10},
    year = {2015},
	langid = {english},
}

@article{theron_narrow_2015,
	title = {Narrow linewidth single laser source system for onboard atom interferometry},
	volume = {118},
	issn = {1432-0649},
	url = {https://doi.org/10.1007/s00340-014-5975-y},
	doi = {10.1007/s00340-014-5975-y},
	pages = {1--5},
	number = {1},
	 journal = {Applied Physics B},
	shortjournal = {Appl. Phys. B},
	author = {Theron, Fabien and Carraz, Olivier and Renon, Geoffrey and Zahzam, Nassim and Bidel, Yannick and Cadoret, Malo and Bresson, Alexandre},
	urldate = {2021-06-08},
	date = {2015-01-01},
    year = {2015},
	langid = {english},
}

@article{wang_agile_2022,
	title = {Agile offset frequency locking for single-frequency fiber lasers},
	volume = {93},
	issn = {0034-6748},
	url = {https://doi.org/10.1063/5.0089303},
	doi = {10.1063/5.0089303},
	pages = {083002},
	number = {8},
	 journal = {Review of Scientific Instruments},
	shortjournal = {Rev. Sci. Instrum.},
	author = {Wang, Enlong and Wang, Guochao and Yu, Xiao and Ying, Kang and Yang, Mingyue and Zhang, Xu and Li, Xuan and Yan, Shuhua and Yang, Jun and Zhu, Lingxiao},
	urldate = {2025-10-28},
	date = {2022-08-19},
    year = {2022},
}

@article{loh_microresonator_2016,
	title = {Microresonator Brillouin laser stabilization using a microfabricated rubidium cell},
	volume = {24},
	issn = {1094-4087},
	url = {https://www.osapublishing.org/abstract.cfm?URI=oe-24-13-14513},
	doi = {10.1364/OE.24.014513},
	pages = {14513},
	number = {13},
	 journal = {Optics Express},
	shortjournal = {Opt. Express},
	author = {Loh, William and Hummon, Matthew T. and Leopardi, Holly F. and Fortier, Tara M. and Quinlan, Frank and Kitching, John and Papp, Scott B. and Diddams, Scott A.},
	urldate = {2021-01-18},
	date = {2016-06-27},
    year = {2016},
	langid = {english},
}

@article{chen_planar-integrated_2022,
	title = {Planar-{Integrated} {Magneto}-{Optical} {Trap}},
	volume = {17},
	issn = {2331-7019},
	url = {https://link.aps.org/doi/10.1103/PhysRevApplied.17.034031},
	doi = {10.1103/PhysRevApplied.17.034031},
	language = {en},
	number = {3},
	urldate = {2025-11-17},
	journal = {Physical Review Applied},
	author = {Chen, Liang and Huang, Chang-Jiang and Xu, Xin-Biao and Zhang, Yi-Chen and Ma, Dong-Qi and Lu, Zheng-Tian and Wang, Zhu-Bo and Chen, Guang-Jie and Zhang, Ji-Zhe and Tang, Hong X. and Dong, Chun-Hua and Liu, Wen and Xiang, Guo-Yong and Guo, Guang-Can and Zou, Chang-Ling},
	month = mar,
	year = {2022},
	pages = {034031},
}

@techreport{steck_rubidium_2001,
	title = {Rubidium 87 {D} {Line} {Data}},
	url = {https://steck.us/alkalidata/rubidium87numbers.1.6.pdf},
	urldate = {2025-11-17},
	institution = {Los Alamos National Laboratory},
	author = {Steck, Danial A},
	month = sep,
	year = {2001},
}

@techreport{thorlabs_collimator,
    author = {{Thorlabs, Inc.}},
    institution = {Thorlabs},
    year = {2021},
	title = {{F810APC}-780-{AutoCADPDF}},
	url = {https://www.thorlabs.com/drawings/6eb3bcb4bdcdce79-C36D5BFC-9849-9B12-5A99830D819EB75B/F810APC-780-AutoCADPDF.pdf},
    howpublished = {Technical data sheet},
}

@manual{Thorlabs_CS135MU_Manual_2025,
  title        = {Compact Scientific Digital Cameras — User Guide (CS Series)},
  author       = {{Thorlabs, Inc.}},
  year         = {2025},
  month        = jul # "~21",
  url          = {https://www.thorlabs.com/drawings/6eb3bcb4bdcdce79-C36D5BFC-9849-9B12-5A99830D819EB75B/CS135MU-Manual.pdf},
  howpublished = {Technical Manual},
}

@article{wiegand_single-laser_2019,
	title = {A single-laser alternating-frequency magneto-optical trap},
	volume = {90},
	issn = {0034-6748, 1089-7623},
	url = {https://pubs.aip.org/rsi/article/90/10/103202/360725/A-single-laser-alternating-frequency-magneto},
	doi = {10.1063/1.5110722},
	language = {en},
	number = {10},
	urldate = {2026-01-05},
	journal = {Review of Scientific Instruments},
	author = {Wiegand, B. and Leykauf, B. and Döringshoff, K. and Gupta, Y. D. and Peters, A. and Krutzik, M.},
	month = oct,
	year = {2019},
	pages = {103202}
}

@techreport{SCIL_NOIP1SN1300A_D_datasheet_2021,
  title        = {NOIP1SN1300A/D -- PYTHON 1.3/0.5/0.3 Megapixels Global Shutter CMOS Image Sensors: Data Sheet},
  author       = {{Semiconductor Components Industries, LLC}},
  institution  = {{ON Semiconductor}},
  address      = {Phoenix, Arizona, USA},
  year         = {2021},
  month        = aug,
  number       = {NOIP1SN1300A/D},
  type         = {Technical data sheet},
  revision     = {Rev. 6},
  url          = {https://www.onsemi.com/download/data-sheet/pdf/noip1sn1300a-d.pdf}
}

@article{lett_observation_1988,
	title = {Observation of Atoms Laser Cooled below the Doppler Limit},
	volume = {61},
	issn = {0031-9007},
	url = {https://link.aps.org/doi/10.1103/PhysRevLett.61.169},
	doi = {10.1103/PhysRevLett.61.169},
	pages = {169--172},
	number = {2},
	 journal = {Physical Review Letters},
	shortjournal = {Phys. Rev. Lett.},
	author = {Lett, Paul D. and Watts, Richard N. and Westbrook, Christoph I. and Phillips, William D. and Gould, Phillip L. and Metcalf, Harold J.},
	urldate = {2022-05-23},
	date = {1988-07-11},
    year = {1988},
	langid = {english},
}

@article{wiegand_linien_2022,
	title = {Linien: A versatile, user-friendly, open-source {FPGA}-based tool for frequency stabilization and spectroscopy parameter optimization},
	volume = {93},
	issn = {0034-6748},
	url = {https://doi.org/10.1063/5.0090384},
	doi = {10.1063/5.0090384},
	shorttitle = {Linien},
	pages = {063001},
	number = {6},
	 journal = {Review of Scientific Instruments},
	shortjournal = {Review of Scientific Instruments},
	author = {Wiegand, B. and Leykauf, B. and Jördens, R. and Krutzik, M.},
	urldate = {2023-11-28},
	date = {2022-06-29},
    year = {2022},
}
\bibliographystyle{sciencemag}

% After the paper has completed peer review and been revised ready for acceptance,
% you should comment out the lines above and copy-paste the contents of your .bbl
% file here instead. This will help ensure that our conversion software works correctly.
% Remember to re-run BibTeX first - check the timestamp!
%
%
%\begin{thebibliography}{1}
%
%\bibitem{example}
%A.~N. {Author}, An example reference. \emph{Journal of Improbable Research}
%  \textbf{1}, 67 (2020).
%\bibitem{example2}
%F.~M. {Surname}, S.~{Author}, A second example. \emph{Interesting Research
%  Letters} \textbf{32}, 897 (2019).
%
%\bibitem{example_preprint}
%P.~{One}, P.~{Two}, P.~{Three}, {An unpublished preprint}. \emph{preprint}
%  (2021), arXiv:2101.12345.
%
%\end{thebibliography}

%%%%%%%%%%%%%%%% ACKNOWLEDGEMENTS %%%%%%%%%%%%%%%

\section*{Acknowledgments}

We thank Catie LeDesma and Kendall Mehling (CU Boulder) and Jeff Pulskamp (Army Research Laboratory) for useful discussions. 

\paragraph*{Funding:}
Research was sponsored by the Army Research Laboratory with SEMI-PNT and was accomplished under Cooperative Agreement Number W911NF-22-2-0050. The views and conclusions contained in this document are those of the authors and should not be interpreted as representing the official policies, either expressed or implied, of the Army Research Laboratory or the U.S. Government or SEMI. The U.S. Government is authorized to reproduce and distribute reprints for Government purposes notwithstanding any copyright notation herein. This research was also sponsored by the Army Research Laboratory Cooperative Agreement Number W911NF-22-2-0056 for PZT fabrication, NSF QuSeC-TAQS (2326784) for PZT integration with atoms and cold-atom experimentation, and NASA Quantum Pathways Institute (80NSSC23K1343) for device characterization. S.C. was supported by the U.S. Department of Energy / National Nuclear Security Administration under Award Number DE-NA0004196. Any opinions, findings, conclusions or recommendations expressed in this material are those of the author(s) and do not necessarily reflect the views of the National Aeronautics and Space Administration (NASA).

\paragraph*{Author contributions:}
A.I., S.C., J.T.C., and D.J.B. prepared the manuscript. A.I. and N.M. set up the PZT PIC devices and did the laser locking. S.C., A.I., and N.M. set up the lock loop for Rb spectroscopy referencing. S.C. ran the MOT temperature characterization. A.I., N.M., and J.W. characterized the PZT PIC devices. J.W. and N.C. designed the devices. N.M. packaged the PZT PIC devices. M.D, with the help of P.M., measured the frequency counter beat-note. S.C. and X.Y. built the Rb MOT experiment. N.I. helped with the MOT temperature characterization and locking electronics. M.W.H. assisted initial work with locked laser tuning and beat-note measurements. C.Z. diced the devices for packaging. I.M.K. fabricated the PZT actuators which were designed by R.Q.R. J.T.C. and D.J.B. supervised and led the scientific collaboration. All authors read and approved the final manuscript.

\paragraph*{Competing interests:}
The work of D.J.B. was funded by ColdQuanta d.b.a. Infleqtion. D.J.B. has consulted for Infleqtion, received compensation, and owns stock. For all other authors, there are no competing interests to declare.

\paragraph*{Data and materials availability:}
Data underlying the results presented in this paper are not publicly available at this time but may be obtained from the authors upon reasonable request.

%%%%%%%%%%%%%%%% SUPPLEMENT LIST %%%%%%%%%%%%%%%

% List the contents of your Supplementary Materials, including the numbers of any
% supplementary figures, tables, external data files etc. and any references that are
% cited only in the supplement. In this example, refs. 7-8 are cited only in the supplement.
% Fill out your numbers accordingly and delete any lines that aren't applicable.
\newpage
\subsection*{Supplementary materials}
Supplementary Text sections S1 and S2\\
Figs. S1 to S6\\
Tables S1 to S2\\
References \textit{(7-\arabic{enumiv})}\\ % automatically fills out the last reference number
% (filling out the other numbers automatically is possible but fiddly and liable to break)

%%%%%%%%%%%%%%%% END OF MAIN TEXT %%%%%%%%%%%%%%%

\newpage

%%%%%%%%%%%%%%%% START OF SUPPLEMENT %%%%%%%%%%%%%%%

% Figures, tables, equations and pages in the supplement are numbered S1, S2 etc.
\renewcommand{\thefigure}{S\arabic{figure}}
\renewcommand{\thetable}{S\arabic{table}}
\renewcommand{\theequation}{S\arabic{equation}}
\renewcommand{\thepage}{S\arabic{page}}
\setcounter{figure}{0}
\setcounter{table}{0}
\setcounter{equation}{0}
\setcounter{page}{1} % not 0 as \newpage already started a supplementary page
% References continue the numbering from the main text.

%%%%%%%%%%%%%%%% SUPPLEMENT TITLE PAGE %%%%%%%%%%%%%%%

\begin{center}
\section*{Supplementary Materials for\\ \scititle}

% Author list for the supplement
% Indicate the corresponding authors, but do NOT include institutions here
% It would be nice if the template auto-generated this, but doing so is complicated...

Andrei~Isichenko and Steven~Carpenter \textit{et al.}

\small
Corresponding authors: \\ Jennifer T. Choy (\texttt{jennifer.choy@wisc.edu}) and Daniel J. Blumenthal (\texttt{danb@ucsb.edu})\\

\end{center}

% Fill out the numbers for each type of supplementary material,
% and delete any lines that aren't applicable.
% These are just example numbers that don't match the rest of this template.
\subsubsection*{This PDF file includes:}
Supplementary Text\\
Figs. S1 to S6\\
Tables S1 to S2\\
References \textit{(7-\arabic{enumiv})}\\ % automatically fills out the last reference number
% (filling out the other numbers automatically is possible but fiddly and liable to break)

\newpage

%%%%%%%%%%%%%%%% MATERIALS AND METHODS %%%%%%%%%%%%%%%

% \subsection*{Materials and Methods}

% \subsubsection*{Example supplement heading}

% The two main sections of the supplement can be split up using headings.

%%%%%%%%%%%%%%%% SUPPLEMENTARY TEXT %%%%%%%%%%%%%%%
\section{PZT resonator characterization and operation}

\subsection*{Small signal modulation and hysteresis}

\begin{figure}[H]
    \centering
	\includegraphics[width=1.0\textwidth]{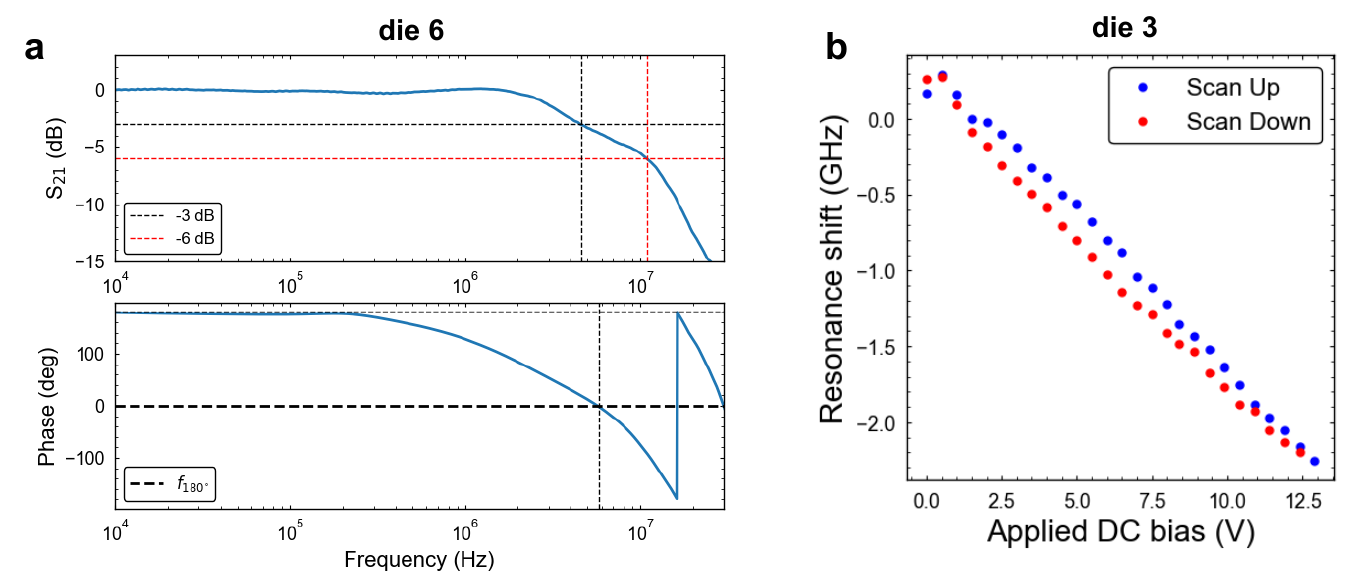} % for an image file named 
	\caption{\textbf{PZT device characterization.} \textbf{a)} Frequency response measurements of device die 6 which had the strongest tuning strength and fastest response. The -3 dB and -6 dB response roll-offs are at 4.5 MHz and 11 MHz, respectively. The 180 degree phase lag point is at 5.8 MHz. \textbf{b)} Hysteresis measurements of the die 3 used for the sub-Doppler demonstration. 
    }
	\label{fig:sup_PZT_AC_hyst}
\end{figure}

We characterized multiple PZT-on-SiN ring resonator devices from a single wafer, and here summarize representative results for two devices: die 6, which exhibited the strongest tuning efficiency, and die 3, which was used for the sub-Doppler cooling demonstration (Fig. \ref{fig:sup_PZT_AC_hyst}). We measured the small signal electrical-to-optical response $S_{21}$ with a Keysight N5247B PNA-X vector network analyzer connected to the PZT actuator electrodes and a fast photodetector. By tuning the laser to the quadrature point of the optical resonance and biasing the PZT at 3 V, we measured the amplitude-modulation frequency response of the PZT-tuned resonator.

For die 6, the -3 dB and -6 dB response roll-offs occur at 4.5 MHz and 11 MHz, respectively, and the $180^{\circ}$ phase-lag point is observed at 5.8 MHz (Fig.~\ref{fig:sup_PZT_AC_hyst}A. For die 3, the -6 dB roll-off is measured at 12.3 MHz. The average electrical resistance and capacitance of these PZT ring resonator devices are approximately 2 $\Omega$ and 700 pF, respectively, corresponding to an RC-limited electrical bandwidth of $f_e = 1/(2\pi RC) \approx 114~\mathrm{MHz}$. The bandwidth limit imposed by the photon lifetime of the cavity is $f_o = c/(\lambda Q_L) = 170~\mathrm{MHz}$ for die 6. The total modulation bandwidth can be expressed as the combined contribution of the electrical, opto-mechanical (OM), and optical limits:
\begin{equation}
\frac{1}{f^2_{\mathrm{BW}}} = \frac{1}{f^2_e} + \frac{1}{f^2_{\mathrm{OM}}} + \frac{1}{f^2_{o}}  
\end{equation}
This analysis indicates that the measured bandwidth of approximately 11 MHz is limited by the opto-mechanical response of the piezoelectric actuator, rather than by electrical or optical constraints.

The PZT actuator exhibits hysteresis which can affect the absolute frequency of the locked laser after the spectroscopy referencing step. Figure~\ref{fig:sup_PZT_AC_hyst}B shows static tuning measurements of die~3, the device used in the sub-Doppler cooling demonstration, where forward (scan-up) and reverse (scan-down) DC voltage sweeps reveal a small bias-dependent hysteresis. This hysteresis manifests primarily as an offset between the two tuning curves, while the tuning slope remains approximately linear over the full voltage range. This results in a history-dependent mapping between applied voltage and laser frequency. During repeated actuation cycles, this effect can lead to slow variations in the cooling laser detuning, which in turn produces observable changes in MOT fluorescence as discussed below. 

\subsection*{Laser locking}

\begin{figure}[H]
	\centering
	\includegraphics[width=0.95\textwidth]{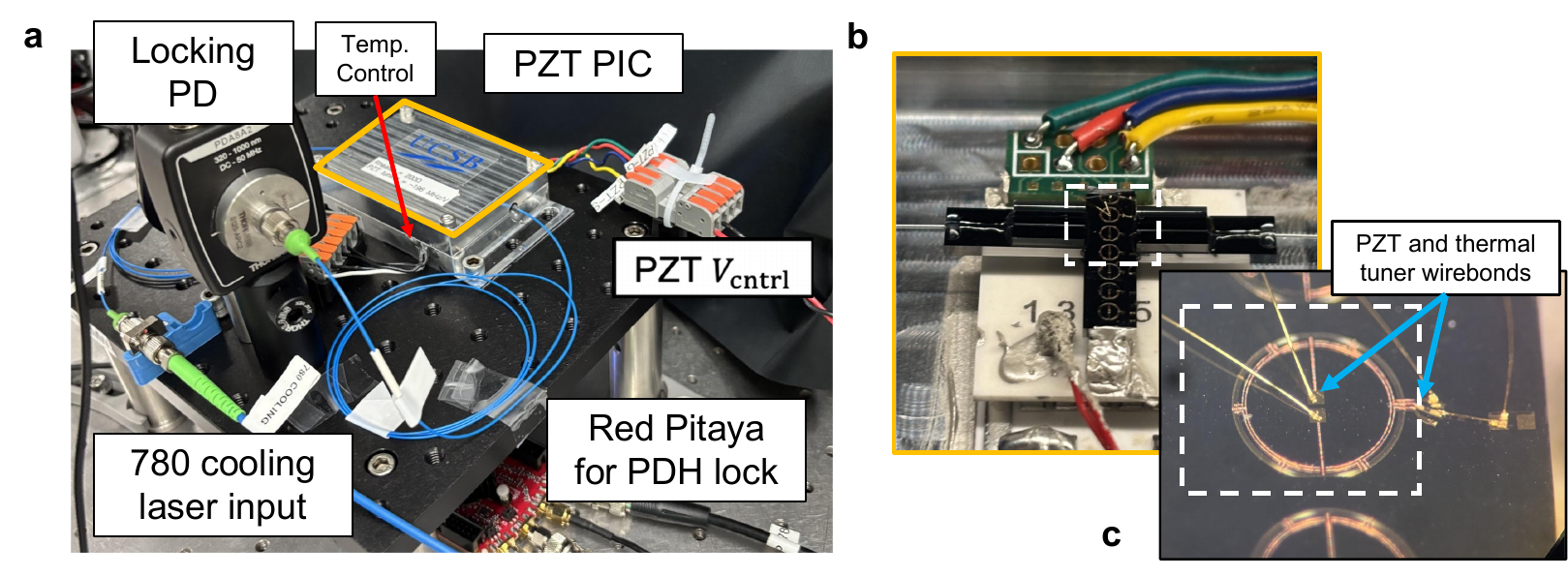} % for an image file named 
	\caption{\textbf{PZT PIC packaging.} \textbf{a)} Setup for laser stabilization and PZT PIC control, with the packaged PIC inside the enclosure highlighted in orange. The package is temperature controlled and has inputs for the PZT control. The Red Pitaya FPGA is used as the servo for the 780 laser lock to the PZT PIC resonator. \textbf{b)} PZT PIC resonator die with v-groove fiber attachments. The die is bonded to a TEC. \textbf{c)} Wirebonded resonator for control of the PZT actuator and thermal tuners. 
}
	\label{fig:PZT_packaging}
\end{figure}

\begin{figure}[H] 
	\centering
	\includegraphics[width=1.0\textwidth]{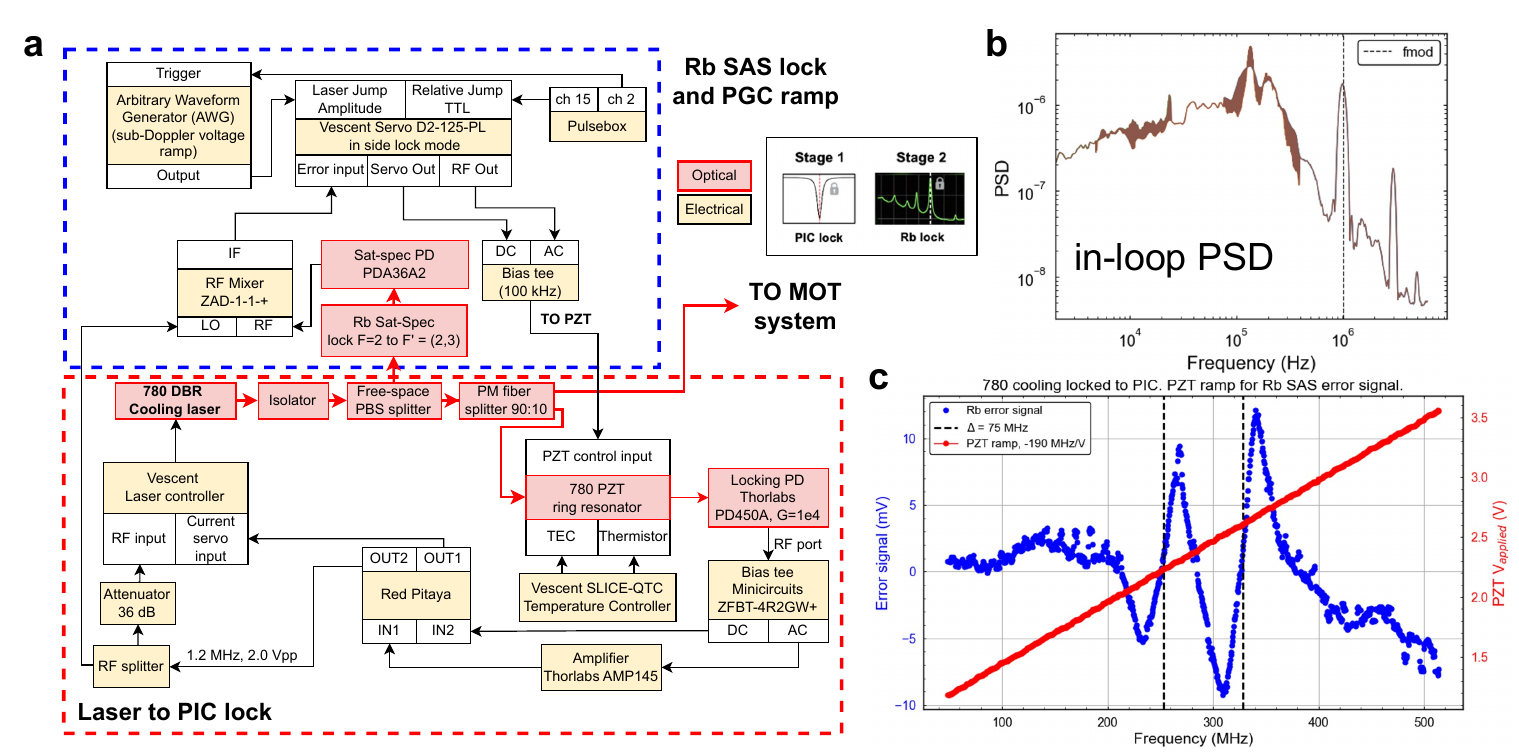} % for an image file named 
	\caption{\textbf{Laser stabilization.} \textbf{a)} Schematic for the laser to PIC locking and the PZT lock to Rb spectroscopy and PGC ramping. \textbf{b)} In-loop power spectral density (PSD) for the laser-to-PIC PDH lock. We use a direct-laser-current 1.2 MHz modulation frequency for generating the error signals for the laser lock and Rb locks. \textbf{c)} Error signal for the Rb spectroscopy lock while ramping the PZT with the laser-to-PIC lock engaged. From the known separation of the hyperfine peaks ($F'=(1,3)$ and $F'=(2,3)$, 75 MHz) we extract a frequency tuning strength of -190 MHz/V which is similar to the measured value in Fig. \ref{fig:res_characterization}.
}
	\label{fig:PZT_laser_locking}
\end{figure}

The PZT-on-SiN resonator device is packaged and integrated into the experimental system as shown in Fig. \ref{fig:PZT_packaging}. The packaged PIC is mounted on a temperature-controlled stage and wirebonded to provide electrical access to both the PZT actuator and integrated thermal tuners (Fig. \ref{fig:PZT_packaging}B,C). Laser stabilization is implemented using a Pound–Drever–Hall (PDH) locking scheme and the schematic of the full laser stabilization architecture is shown in Fig. \ref{fig:PZT_laser_locking}A. The PDH servo for the laser to PIC lock is implemented on a Red Pitaya FPGA platform using the Linien locking application \cite{wiegand_linien_2022} and the Rb spectroscopy lock uses a Vescent D2-125-PL servo for its triggerable, relative current jump feature. To generate error signals for both locking stages, we apply a single 1.2-MHz modulation tone to the 780-nm DBR laser via direct current modulation. We attenuate the signal into the laser current controller to limit the frequency modulation span and any added noise. This tone is used as the LO demodulation for both locks. The resulting in-loop frequency noise power spectral density (PSD), extracted from the PDH error signal, is shown in Fig. \ref{fig:PZT_laser_locking}B. With the laser-to-PIC lock engaged, frequency tuning of the locked laser is achieved by applying a voltage to the PZT actuator. Figure~\ref{fig:PZT_laser_locking}C shows the error signal obtained while linearly ramping the PZT voltage and scanning across the rubidium saturation-absorption spectroscopy (SAS) features. 


\section{MOT system and characterization and measurement}
\label{MOT system }

We used a six-beam (three beams with retro-reflection) free-space $^{87}$Rb magneto-optical trap (MOT). The 780-nm laser drives a red-detuned $\mathrm{^{87}Rb}~5^{2}S_{1/2}~F=2 \rightarrow 5^{2}P_{3/2}~F^\prime =3$ transition, while a separate 795-nm DBR laser acts as the repump, addressing the $\mathrm{^{87}Rb}~5^{2}S_{1/2}~F=1 \rightarrow 5^{2}P_{1/2}~F^\prime =2$ transition. Our system uses an home-built 2D PCB coil originally inspired by Chen et al. \cite{chen_planar-integrated_2022}. The coils are switched by a custom circuit with $\sim 100 ~\mathrm{\mu s}$ switching time. In all configurations, we use a semiconductor optical amplifier (Thorlabs BOA780P) for cooling power control and shuttering.

\subsection*{System configurations in this work}

\begin{figure}[H]
	\centering
	\includegraphics[width=0.6\textwidth]{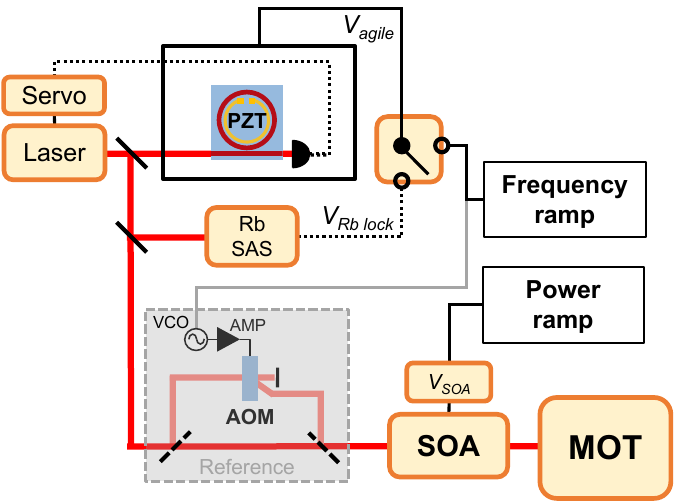} % for an image file named 
	\caption{\textbf{MOT system configurations for temperature measurements. }
	The 780-nm cooling laser is locked to the PZT agile cavity. The laser tracks the resonance frequency of the cavity, which is itself either tied to rubidium saturation-absorption spectroscopy (SAS) or to the applied frequency ramp signal. The power ramp signal modulates the cooling beam power delivered to the MOT via current through the semiconductor optical amplifier (SOA). MOT temperatures were recorded for four cases: (I) the MOT without PGC (i.e. without any frequency or power ramp), (II) PZT-controlled frequency ramp (without any power ramp), (III) PZT-controlled frequency ramp and SOA-controlled power ramp, and (IV) AOM-controlled frequency ramp and SOA-controlled power ramp (AOM reference). For cases I, II, III the AOM is not used. The AOM is powered by an RF amplifier (AMP) and a voltage controlled oscillator (VCO).
}
	\label{fig:sup_schematics_during_TOF}
\end{figure}

We performed sub-Doppler cooling using four hardware and control configurations, summarized in Fig.\ref{fig:results_temperature}A. A reference configuration, labeled Case IV in in Fig. \ref{fig:results_temperature}, employed an AOM (see Fig.\ref{fig:sup_schematics_during_TOF}). In this conventional setup, the laser was locked to the $\mathrm{^{87}Rb},5^2S_{1/2},F=2 \rightarrow 5^2P_{3/2},F^\prime =1,3$ crossover and double-passed through an AOM set near 95 MHz, yielding a red detuning of $\sim$22 MHz from the cooling transition. Both frequency and intensity ramps were applied by modulating the AOM drive and SOA bias, respectively. Figure \ref{fig:results_temperature} shows the TOF data for the coldest Rb cloud achieved in this configuration.

All other configurations (I, II, IV) employ the dual-stage lock (laser-to-cavity and cavity-to-Rb SAS) with the AOM removed from the optical path. In each case, the optical cavity was locked to the $\mathrm{^{87}Rb},5^2S_{1/2},F=2 \rightarrow 5^2P_{3/2},F^\prime=2,3$ crossover (note the distinction from the $F^\prime = 1,3$ crossover of configuration IV) during the Rb-disciplined portion of the experiment (Fig.~\ref{fig:exp_diagram}C). After 223 ms of MOT formation, either a 12 ms polarization-gradient cooling (PGC) ramp sequence was applied (II and III) or the MOT was held constant (I). The atomic cloud temperature is determined from time-of-flight (TOF) imaging (more information below). Two experimentally optimized PGC setups are tested: frequency-only ramping (II), where the laser was further red-detuned via a voltage ramp on the PZT actuator, and combination frequency and intensity ramping using the PZT and SOA, respectively (III). For comparison, Fig.~\ref{fig:results_temperature} includes the AOM-based configuration. The integrated tunable-cavity approach provides a larger and more stable (in optical power) tuning range without affecting beam alignment, offering a compact and power-efficient alternative for sub-Doppler atom cooling.

\subsection*{Atom number calculation}
\label{subsubsec:Atom_number_calc}

MOT fluorescence, in combination with other experimental conditions, can be used to determine the number of trapped atoms:
\begin{equation}
  \label{eq:Atom number}
  N
  \;=\;
  \frac{1}
       {R_{\mathrm{scat}}} \ \dv{N_{\mathrm{\gamma}}}{t},
\end{equation}
where $\dv{N_{\mathrm{\gamma}}}{t}$ is the rate of photons emitted over the entire sphere in a unit time. The photon scattering rate $R_{\mathrm{scat}}$ is given by \cite{steck_rubidium_2001}
\begin{equation}
  \label{eq:Rscat}
  R_{\mathrm{scat}}
  \;=\;
  \frac{\Gamma}{2}\,
  \frac{\frac{I_{\mathrm{tot}}}{I_{\mathrm{sat}}}}
       {1 + \frac{I_{\mathrm{tot}}}{I_{\mathrm{sat}}} + 4\left(\frac{\Delta}{\Gamma}\right)^2\,}
\end{equation}
where $\Delta = 2\pi\,\delta f_{\mathrm{MHz}}$, and $\delta f_{\mathrm{MHz}}$ is the frequency detuning in MHz of the cooling laser with respect to the $^{87}$Rb $5^2S_{1/2}\,F=2 \rightarrow 5^2P_{3/2}\,F' = 3$ transition. $I_{tot}$ and $I_{sat}$ are the total cooling beam intensity and the saturation energy, respectively. We used a plano-convex lens just outside our glass vacuum chamber to collect light. This light was imaged with a traditional compound camera lens onto the sensor of our Thorlabs CS135MU camera.

From the digital images captured by the camera, we can determine $\dv{N_{\mathrm{\gamma}}}{t}$ by summing the (background-subtracted) pixel counts in our region of interest ($C_{x,y}$) from an exposure and applying conversion factors:
\begin{equation}
    \label{eq: MOT optical power}
    \dv{N_{\mathrm{\gamma}}}{t} = 
    \frac{ 4 \pi
        \left( \sum_{ROI_{x,y}} C_{x,y} \right) G}
        {\Omega~\eta_{\mathrm{QM}}~\eta_{\mathrm{geo}}~\Delta t_{exp}},
\end{equation}
where $G$ is the gain coefficient of the camera sensor, $\eta_{QM}$ is the quantum efficiency of the camera at 780 nm, $\eta_{geo}$ accounts for a geometric loss in the light collection cone from an obstruction in the imaging path, $\Omega$ is the solid angle of light collection, and $\Delta t_{exp}$ is the duration of camera exposure. The values for $\Omega$ and $\eta_{geo}$ were found by careful measurement of the setup and simulation in Zemax OpticStudio$^{\mathrm{TM}}$. Table \ref{tab:sup_imaging_params} shows the experimental parameters used in the calculation of the atom number.

When capturing MOT images just before PGC (at $\tau_{\mathrm{seq}} = 237 ~\mathrm{ms}$ as seen in Fig. \ref{fig:sup_timing_diagram}), we determined an atom number of $\sim5$ million. However, the MOT is loaded for only 227 ms, which is insufficient for full reloading (despite partial atom recapture) between cycles. Longer loading times would increase the atom number but reduce the achievable sampling rate in a full quantum sensor.

% Supplementary Table 1
\setlength{\tabcolsep}{6pt}
\renewcommand{\arraystretch}{1.12}
\small
\begin{longtable}{%
  | >{\centering\arraybackslash}p{0.12\textwidth}
  | >{\raggedright\arraybackslash}p{0.48\textwidth}
  | >{\raggedright\arraybackslash}p{0.22\textwidth}
  | >{\centering\arraybackslash}p{0.05\textwidth} |
}
\caption{Experimental parameters used in the determination of MOT atom number.}
\label{tab:sup_imaging_params} \\
\hline
\textbf{Parameter} & \textbf{Explanation} & \textbf{Value} & \textbf{Ref.} \\ \hline
\endfirsthead

\hline
\textbf{Parameter} & \textbf{Explanation} & \textbf{Value} & \textbf{Ref.} \\ \hline
\endhead
% --- Fundamental atomic parameters ---------------------------------
$\lambda$
  & $^{87}$Rb D2 transition wavelength
  & 780.241 nm
  & \cite{steck_rubidium_2001}
  \\\hline
$\Gamma$
  & $^{87}$Rb natural linewidth for the $\mathrm{D}_2$ $(5^2S_{1/2} \rightarrow 5^2P_{3/2})$ transition
  & $2\pi\,(6.066~\mathrm{MHz})$
  & \cite{steck_rubidium_2001}
  \\\hline
$\sigma_0$
  & $^{87}$Rb resonant cross section intensity for the $F=2 \rightarrow F'=3$ and isotropic light polarization
  & $1.36\times 10^{-12}\ \mathrm{cm}^2$
  & \cite{steck_rubidium_2001}
  \\\hline
$I_{\mathrm{sat}}$
  & $^{87}$Rb saturation intensity for the $F=2 \rightarrow F'=3$ transition with isotropic light polarization
  & $3.577~\mathrm{mW\,cm^{-2}}$
  & \cite{steck_rubidium_2001}
  \\\hline
% --- MOT/beam parameters -------------------------------------------
$\Delta$
  & (Angular) frequency detuning from the $F=2 \rightarrow F'=3$ transition (for configurations I, II, and III). Determined by beat-note measurements
  & $-2\pi\,(25.5 \pm 0.2~\mathrm{MHz})$
  &
  \\\hline
$P_{\mathrm{x, y}}$
  & Average power of the x- and y-axis cooling beams measured just before entering the cell
  & $6.22 \pm 0.05$ mW
  &
  \\\hline
$P_{\mathrm{z}}$
  & Average power of the z-axis beam measured just before entering the cell. Since this branch goes through a beam expander, and is cut off by the aperture of the PCB coils (6.5 mm diameter), this value is measured after the aperture
  & $1.00 \pm 0.05$ mW
  &
  \\\hline
$A_{\mathrm{x, y}}$
  & Area of the x- and y-axis cooling beams, using the $1/e^2$ beam width of the Thorlabs F810APC-780 collimator
  & $0.442~\mathrm{cm}^2$
  & \cite{thorlabs_collimator}
  \\\hline
$A_{\mathrm{z}}$
  & Area of z-axis cooling beam, using the width of the PCB coil aperture (6.5 mm diameter)
  & $0.332~\mathrm{cm}^2$
  &
  \\\hline
$I_{\mathrm{x, y}}$
  & Average intensity of x- and y-axis cooling beams before the glass cell
  & $P_{\mathrm{x, y}}/A_{\mathrm{x, y}} =  14.07 \pm 0.11 ~\mathrm{mW\,cm^{-2}}$
  &
  \\\hline
$I_{\mathrm{z}}$
  & Average intensity of z-axis cooling beam
  & $P_{\mathrm{z}}/A_{\mathrm{z}} =  3.00 \pm 0.15 ~\mathrm{mW\,cm^{-2}}$
  &
  \\\hline
$T_{\mathrm{cell}}$
  & Measured transmission through a layer of the uncoated glass vacuum cell at $45^\circ$ incidence
  & $0.96 \pm 0.01$
  &
  \\\hline
$I_{\mathrm{tot}}$
  & Sum of the intensities of the six trapping beams at the location of atoms. Three input beams and three retro-reflected (RR) beams. The z-axis beam passes through the AR cell at normal incidence with negligible reflections. Assume perfect retro-reflection
  & $2T_{\mathrm{cell}}\,I_{\mathrm{x, y}} + 2T_{\mathrm{cell}}^2\,I_{\mathrm{x, y}} + 2I_{\mathrm{z}} = 59.0 \pm 1.0 ~\mathrm{mW\,cm^{-2}}$
  &
  \\\hline
% --- Collection and detection parameters ---------------------------
$\Omega$
  & Solid angle of collected light. This value was calculated from experimental measurements and lens parameters simulated in Zemax OpticStudio. The resulting half-angle collection cone was $6.0^\circ$
  & $0.035$ sr
  &
  \\\hline
$\eta_{geo}$
  & Geometrical light collection efficiency, assuming the transmission through the B-coated lenses are nearly 1 (this light is collected at $\leq 6^\circ$ from normal incidence). This loss is the result of slight clipping of the collection cone by an optical mount
  & $\eta_{geo} = 0.74$
  &
  \\\hline
$\eta_{QM}$
  & Camera (Thorlabs CS135MU) quantum efficiency at $\lambda$ = 780 nm
  & $0.45~\mathrm{e^{-}}$ per photon
  & \cite{Thorlabs_CS135MU_Manual_2025}
  \\\hline
$G$
  & Camera sensor (onsemi NOIP1SN1300A-QTI) gain coefficient. LSB means a single ``Least-Significant Bit'' flip
  & $10.42~\mathrm{e^{-}}$ per LSB
  & \cite{SCIL_NOIP1SN1300A_D_datasheet_2021}
  \\\hline

$\Delta t_{exp}$
  & Camera exposure time
  & 0.5 ms (single-shot), 5 ms (10x accumulation)
  & 
  \\\hline
\end{longtable}

\subsection*{Temperature measurement}
\label{subsubsec:Temp_data_collection}

\begin{figure}[H]
    \centering
	\includegraphics[width=0.85\textwidth]{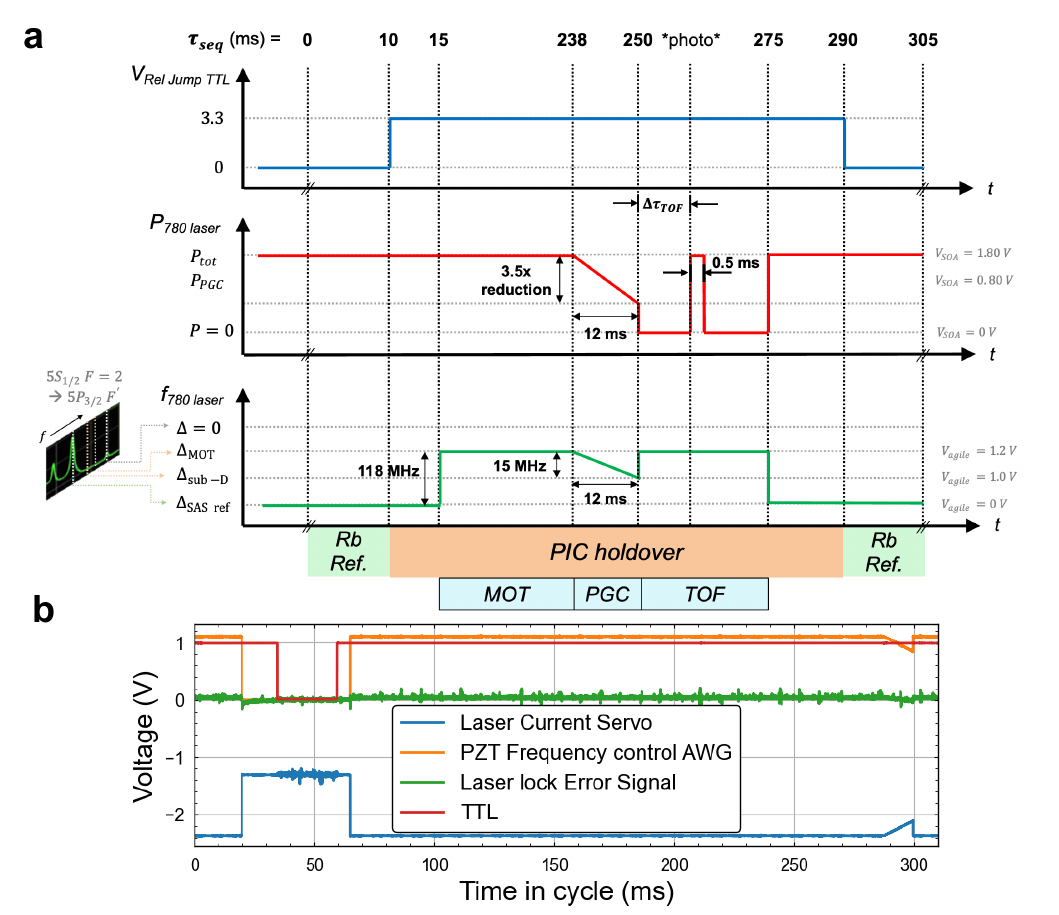} % for an image file named 
	\caption{\textbf{Experiment timing.}
    \textbf{a)} Timing diagram for configuration III of Fig.~\ref{fig:results_temperature} and Fig.~\ref{fig:sup_schematics_during_TOF}. All time-of-flight (TOF) measurements are performed within a repeating 305 ms control sequence, with the TOF delay $\tau_{\mathrm{TOF}}$ varied between runs. Saturation-absorption spectroscopy (SAS) re-referencing occurs during two 25 ms windows at the beginning and end of each cycle when the relay-jump TTL is low. MOT loading begins at $\tau_{\mathrm{seq}} = 10$ ms and continues until 238 ms for sequences including polarization-gradient cooling (PGC), or until 250 ms when PGC is omitted. During PGC, the magnetic field is switched off at 238 ms (250 ms without PGC) and remains off until 275 ms. At $\tau_{\mathrm{seq}} = 250$ ms, the cooling and repump beams are shuttered and the atomic cloud undergoes ballistic expansion for a variable $\tau_{\mathrm{TOF}}$. Imaging is performed using a 0.5 ms camera exposure, after which the laser frequency is returned to the SAS reference point to prepare for the next cycle. \textbf{b)} Monitoring signals during the sequence, showing the lock error signal (green), which confirms that the laser remains locked to the PZT-stabilized cavity throughout the cycle.
}
	\label{fig:sup_timing_diagram}
\end{figure}

The temperature of our $^{87}$Rb MOT is determined by imaging the ballistically expanding atom cloud with a 0.5 ms exposure for a series of different $\tau_{\mathrm{TOF}}$ (time-of-flights). At $\tau_{\mathrm{TOF}} = 0$, we shutter both cooling and repump lasers and rapidly shut off the magnetic coils. Note that we use an earlier coil shutoff for polarization-gradient cooling. Timing information for our experimental determination of the MOT temperature is shown in Fig. \ref{fig:sup_timing_diagram}. 

For six distinct $\tau_{\mathrm{TOF}}$ settings (1, 2, 3, 4, 5, 6 ms), we collected background-subtracted images and fit the cloud widths to the linearized ballistic expansion equations:
\begin{equation}
  \label{eq:TOFs}
  \sigma_y^2(\tau_{\mathrm{TOF}}) \;=\; \sigma_{y,0}^2 + \frac{k_{\mathrm{B}}T_y}{m}\,\tau_{\mathrm{TOF}}^2,
  \qquad
  \sigma_z^2(\tau_{\mathrm{TOF}}) \;=\; \sigma_{z,0}^2 + \frac{k_{\mathrm{B}}T_z}{m}\,\tau_{\mathrm{TOF}}^2,
\end{equation}
where $\sigma_y(\tau_{\mathrm{TOF}})$, $\sigma_z(\tau_{\mathrm{TOF}})$ are the standard deviations of  Gaussian fits to the cloud images in each dimension, $l_{\mathrm{px}}$ is the calibrated pixel size, $k_B$ is the Boltzmann constant, $m$ is the mass of $^{87}$Rb, and $T_y$,$T_z$ are the MOT temperatures along the $y$,$z$ axes. Table \ref{tab:sup_temp_params} lists the relevant values for this calculation.

% Supplementary Table 1
\setlength{\tabcolsep}{6pt}
\renewcommand{\arraystretch}{1.12}
\small
\begin{longtable}{%
  | >{\centering\arraybackslash}p{0.12\textwidth}
  | >{\raggedright\arraybackslash}p{0.40\textwidth}
  | >{\raggedright\arraybackslash}p{0.30\textwidth}
  | >{\centering\arraybackslash}p{0.05\textwidth} |
}
\caption{Parameters used in the determination of atom temperatures}
\label{tab:sup_temp_params} \\
\hline
\textbf{Parameter} & \textbf{Explanation} & \textbf{Value} & \textbf{Ref.} \\ \hline
\endfirsthead

\hline
\textbf{Parameter} & \textbf{Explanation} & \textbf{Value} & \textbf{Ref.} \\ \hline
\endhead
% --- Fundamental atomic parameters ---------------------------------
$k_B$
  & Boltzmann constant
  & $1.380649*10^{-23} \mathrm{J}~\mathrm{K^{-1}}$
  & 
  \\\hline
$m_{\mathrm{^{87}Rb}}$
  & $^{87}$Rb isotope mass
  & $1.44316060(11)*10^{-25} ~\mathrm{kg}$
  & \cite{steck_rubidium_2001}
  \\\hline
$\Delta t_{exp}$
  & Camera exposure time
  & 0.5 ms (single-shot), 5 ms (10x accumulation)
  & 
    \\\hline
$l_{\mathrm{px}}$
  & USAF target calibrated pixel size
  & $12.86 ~\mathrm{\mu m ~ px^{-1}}$
  & 
  \\\hline
\end{longtable}

\subsection*{Atom-number drift arising from PZT frequency-response drift}
\begin{figure}[H]
    \centering
	\includegraphics[width=1.0\textwidth]{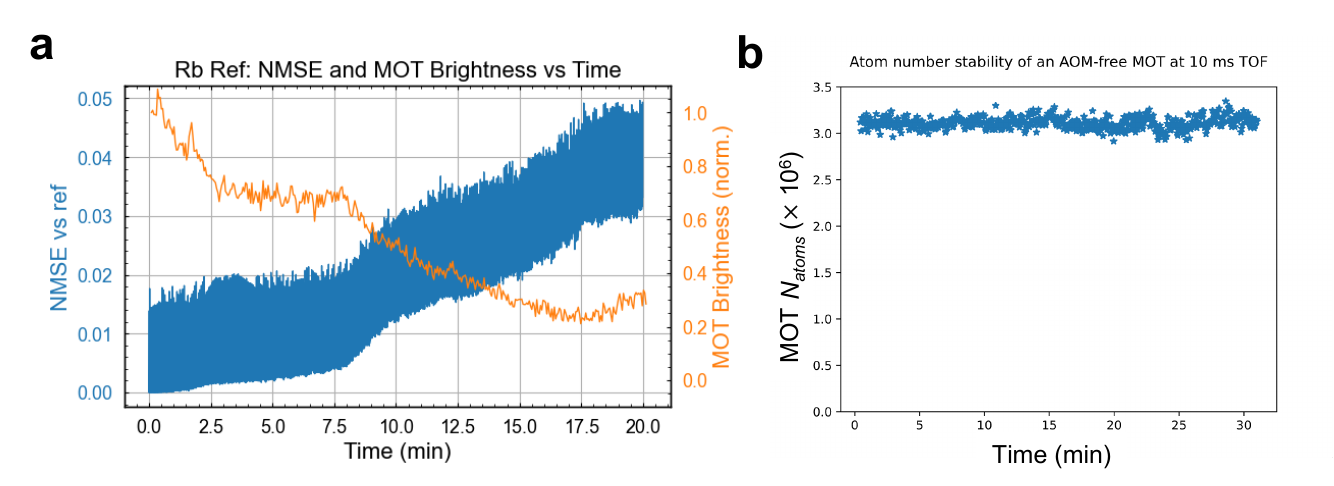} % for an image file named 
	\caption{\textbf{PZT Hysteresis and atom number stability}
    \textbf{a)} Extended time-series data for Fig. \ref{fig:results_beatnote_control_monitor}B. Normalized mean squared error (NMSE) of the beat note between the cooling and reference lasers (blue) and number of atoms in the MOT (orange) measured over the same time interval. The NMSE is calculated between the frequency trace of each sequence w.r.t. the first sequence. The slow upward drift of the NMSE corresponds to changes in the PZT voltage-to-frequency actuation. Variations in the trapped atom number correlate with this drift. In this test, the MOT operates with reduced cooling beam power owing to light diverted for the simultaneous beat-note measurement. \textbf{b)} Continuous 30-minute record of MOT atom number using the PZT cavity for spectroscopy referencing and for PGC, during which the PZT hysteresis was negligible.}
	\label{fig:sup_hysteresis_and_atom_number}
\end{figure}

Figure \ref{fig:results_beatnote_control_monitor}B in the main text shows MOT atom number measurements (red curve) over several minutes using two schemes: either periodic referencing to Rb spectroscopy (upper plot) or halting Rb-referencing at $t=0$ (lower plot). The extended dataset for the former case is shown in Figure \ref{fig:sup_hysteresis_and_atom_number}A, where we simultaneously record the beat-note frequency (blue curve) between the 780 nm DBR cooling laser and a separate 780 nm DBR locked to the $\mathrm{^{87}Rb},~5S_{1/2},F=2 \rightarrow 5P_{3/2},F^\prime =1,3$ crossover. The left axis shows the normalized mean squared error (NMSE) of the beat-note signal for each MOT + PGC sequence with respect to the sequence for $t=0$.

Without Rb-referencing (Fig, \ref{fig:results_beatnote_control_monitor}B, lower), the MOT completely dissolves within 30 seconds due to the drift of the PZT cavity-locked laser. With Rb-referencing, residual frequency excursion can be attributed to the drift in the PZT cavity voltage-to-frequency response at a fixed voltage jump. As shown in Figure \ref{fig:sup_hysteresis_and_atom_number}B, the isolated effect of PZT hysteresis can be observed: the MOT is still visible for over 20 minutes, though temporal shifts in PZT response result in the loss of $\sim$2/3 of trapped atoms over that time. Without Rb-referencing, the cooling laser exhibits a $\sim$50-MHz drift over 20 minutes, whereas with Rb-referencing, the drifts are reduced to the single-MHz range over the same interval. For precision quantum applications such as computing or sensing, such laser frequency drifts must be minimized. To address this, future work will implement dynamic adjustment of the PZT voltage to counteract these drifts, initially using spectroscopy signals and then ultimately using MOT fluorescence itself for feedback.

We note that PZT frequency-response drift is temporally variable, typically strongest shortly after system power-up. The data in Figure \ref{fig:results_beatnote_control_monitor}B was collected during this initial phase. On a separate day, after several hours of continuous operation, the MOT brightness remained stable for 30 minutes (\ref{fig:sup_hysteresis_and_atom_number}B). In this trial, the cavity was periodically referenced to rubidium spectroscopy (as in Figure \ref{fig:results_beatnote_control_monitor}A) and our typical 12 ms power and frequency PGC ramp was performed. The atom number imaging took place after at 10-ms time-of-flight. Although beat note data was not collected for this trial, the stability of atom number directly corresponds to a stable absolute laser frequency, suggesting the anticipated system performance once dynamic PZT control is implemented. 

%%%%%%%%%%% CAPTIONS FOR OTHER SUPPLEMENTARY FILES %%%%%%%%%%

\clearpage % Clear all remaining figures and tables then start a new page

%%%%%%%%%%%%%%%% SUPPLEMENTARY REFERENCES %%%%%%%%%%%%%%%

% Do NOT include a reference list in the supplement.
% All references must be in a single list at the end of the main text.
% The copyeditors will ensure that the correct reference list appears with each version of the paper
% (print, HTML, PDF, mobile app, metadata for bibliographic databases etc.)

\end{document}